\documentclass[conference]{IEEEtran}
\IEEEoverridecommandlockouts

\usepackage{booktabs}  
\usepackage{array}     
\usepackage{caption}   
\usepackage{subcaption} 
\usepackage{url}
\usepackage[hidelinks]{hyperref}
\usepackage{placeins}

\usepackage[linesnumbered,ruled,vlined]{algorithm2e}

\usepackage{cite}
\usepackage{amsmath,amssymb,amsfonts}
\usepackage{algorithmic}
\usepackage{graphicx}
\usepackage{textcomp}
\usepackage{xcolor}
\usepackage{balance}
\newcommand\cparagraph[1]{\vspace{1mm}\noindent\textbf{#1.}\xspace}
\def\BibTeX{{\rm B\kern-.05em{\sc i\kern-.025em b}\kern-.08em
    T\kern-.1667em\lower.7ex\hbox{E}\kern-.125emX}}

\newcommand{\SystemName}{\textsc{MpGEMM}\xspace}
\begin{document}

\title{Demystifying ARM SME to Optimize General Matrix Multiplications}

\renewcommand{\footnoterule}{
  \vspace*{-3pt}
  \hrule width 0.4\columnwidth
  \vspace*{2.6pt}
}

\author{
\IEEEauthorblockN{
Chencheng Deng\textsuperscript{\S}\thanks{{\S}Equal contribution \quad *Corresponding author},
Weiling Yang\textsuperscript{\S},
Jianbin Fang\textsuperscript{*},
Dezun Dong\textsuperscript{*}}
\IEEEauthorblockA{\textit{College of Computer Science and Technology}, 
\textit{National University of Defense Technology}, Changsha, China \\
\{chenchengdeng, w.yang, j.fang, dong\}@nudt.edu.cn}
}

\maketitle
\begin{abstract}
General Matrix Multiplication (GEMM) is a critical kernel in high-performance computing and deep learning. While modern architectures like ARM's Scalable Matrix Extension (SME) introduce dedicated hardware for matrix operations, existing linear algebra libraries fail to fully exploit its potential, particularly for large matrices. This paper presents \SystemName, an open-source library that leverages key architectural features of SME to optimize GEMM across multiple precisions. Through a systematic characterization of SME, we derive optimization guidelines that inform our design. \SystemName employs cache-aware partitioning, efficient data packing with on-the-fly transposition, and specialized micro-kernels that utilize multi-vector loads and all available tile registers. Evaluated on an Apple M4 Pro with real-world workloads from DeepSeek and LLaMA, \SystemName achieves an average speedup of 1.23$\times$ over the vendor-optimized Apple Accelerate library and significantly outperforms other open-source alternatives.


\end{abstract}


\section{Introduction}
General Matrix Multiplication (GEMM) is a key subroutine in high-performance computing (HPC), spanning workloads from traditional scientific simulations to emerging deep learning (DL). Driven by the continuously growing computational demands, hardware vendors have introduced specialized accelerators for matrix operations, such as NVIDIA Tensor Cores~\cite{tensorcore}, Google TPUs~\cite{tpu}, ARM SMEs~\cite{sme}, and Intel AMXs~\cite{amx}.
To fully exploit the performance potential of these accelerators, it is essential to develop GEMM implementations tailored to their underlying architectural features.

ARM processors play an increasingly important role, not only in embedded devices and data centers due to their low power consumption~\cite{libshalom,DBLP:conf/ics/FuYDS24}, but also as a foundation for scalable HPC solutions.
ARM SME is an architecture and ISA extension designed to enhance matrix operations based on SVE~\cite{sve}.
The SME is designed to deliver high throughput for matrix operations through outer product data processing instructions and a dedicated tile-based accumulator (ZA)~\cite{sme}.

Recently, efforts have been made to leverage SME to optimize GEMM, yet existing implementations exhibit  limitations~\cite{accelerate,libxsmm,openblas,kleidiai}. Accelerate~\cite{accelerate} provides a highly optimized BLAS library on SME, but it is closed-source and restricted to Apple platforms. 
LIBXSMM~\cite{libxsmm} is the first open-source library to exploit SME for high-performance GEMM. It generates micro-kernels in SME2 assembly via just-in-time compilation, but its design primarily targets small-scale GEMMs. OpenBLAS~\cite{openblas} adopts handwritten SME kernels to accelerate small GEMMs, but currently lacks support for SME2 multi-vector instructions. KleidiAI~\cite{kleidiai} is a lightweight library that accelerates fundamental operators using ARM CPU vector extensions, including NEON MLA and SME2 outer product.
As we will show later in the paper, these open-source libraries leave much room for improvement in large-scale GEMMs.

We have identified two key performance bottlenecks in state-of-the-art implementations on SME: \textit{inefficient cache utilization} and \textit{underutilized memory bandwidth}.
On the one hand, our microbenchmarks reveal that the SME unit shares the L2 cache with the CPU cores. However, these open-source libraries all adopt simple three-level nested loops designed to satisfy outer product computation. This leads to sub-blocks that are poorly matched to the shared L2 cache, preventing the packed data from fully benefiting from cache locality. Moreover,
LIBXSMM and OpenBLAS always pack only a single input matrix to reduce the memory overhead of packing. When the unpacked matrix exceeds the capacity of the shared L2 cache, it will incur significant cache misses. These large-stride and discontiguous accesses are also challenging for TLB and hardware prefetchers. These behaviors substantially limit the overall performance of GEMM.
On the other hand, 
microbenchmarking reveals that loading data in groups of four scalable vector (Z) registers can achieve nearly 900 GB/s bandwidth, far surpassing the 230 GB/s of single-register loads.
Existing micro-kernels and packing methods fail to leverage this capability, leaving performance memory-bound.

This paper presents \SystemName, 
an open-source library for high-performance, multiple-precision GEMM on ARM SME.
We introduce an analytical model to partition large matrices into smaller submatrices that fit within the shared L2 cache. These submatrices are packed into memory layouts optimized for efficient multi-vector load and outer product instructions, improving memory access patterns.
We employ on-the-fly transposition and first-round online packing strategies to mitigate the memory access overhead of data packing. Guided by microbenchmarks, we develop high-performance GEMM main and edge micro-kernels that always utilize four scalable vector registers per group for data loading and all ZA tiles for accumulation. Tasks are dynamically distributed across the available SME units to achieve parallel computation. These optimizations are extended to mixed-precision GEMMs with lower-precision inputs and higher-precision outputs, which further improve the compute-to-memory ratio.

We evaluate \SystemName on an Apple M4 Pro chip using real-world GEMM workloads from the DeepSeek and LLaMA models. Results show that \SystemName outperforms the vendor-optimized Apple Accelerate by an average of 1.23$\times$ and achieves average speedups of 3.96$\times$, 4.69$\times$, and 5.7$\times$ over LIBXSMM, KleidiAI, and OpenBLAS, respectively.
Moreover, our mixed-precision GEMM implementation reaches up to 94\% of the peak performance measured on SME.

This paper makes the following contributions:
\begin{enumerate}
\item We systematically characterize the architectural properties of SME, providing critical insights for optimizing other kernels (Section~\ref{sec:microbenchmark}).
\item We introduce a cache-aware algorithm design for GEMM. We develop an analytical model to guide matrix partitioning, transforming the naive three-loop structure into a six-level blocked algorithm that optimizes for the shared L2 cache and TLB (Section~\ref{sec:design}).
\item 
 We introduce new packing schemes that leverage on-the-fly transposition and first-round online packing, alongside micro-kernels that use four-Z-register loads and all available ZA tiles, maximizing computational throughput and memory bandwidth (Sections~\ref{sec:design} and~\ref{sec:mpdesign}).
\item 
We demonstrate that our optimizations extend effectively to mixed-precision GEMMs, achieving up to 94\% of SME's peak performance (Section~\ref{sec:mpdesign}).
\end{enumerate}
\section{Background and Motivation}
\subsection{ARM Scalable Matrix Extension (SME)}
ARM SME is a recent architectural and ISA extension that improves matrix computation capabilities. Building upon the Scalable Vector Extension (SVE), including Streaming SVE (SSVE) mode, a two-dimensional ZA storage, load/store instructions from/to the ZA storage, outer product instructions, among others. SME2 further extends these capabilities with multi-vector instructions, enabling simultaneous operations on multiple scalable vector registers. 

SSVE mode typically employs a longer scalable vector length (SVL) than non-streaming mode to improve data processing throughput.
SME operations require switching to SSVE mode, which provides 32 scalable vector registers (Z0–Z31) for data operands and 16 predicate registers (P0–P7 and PN8–PN15) for masking.
SME introduces a dedicated ZA storage, a two-dimensional architectural matrix register of size $SVL \times SVL$, for storing intermediate results.
On our experimental platform, the Apple M4 Pro, the streaming SVL is fixed at 512 bits, making each Z register 64 bytes in size and the ZA storage 64 × 64 = 4096 bytes. 
The ZA storage can be accessed as vectors with element types of 8, 16, 32, 64, or 128 bits. To maximize storage utilization, it can also be accessed as a set of two-dimensional tiles. The number of tiles depends on the element type shown in Table~\ref{tab:za_tile}. For instance, the ZA storage can be viewed as four tiles (ZA0.S-ZA3.S) for 32-bit elements and eight tiles (ZA0.D-ZA7.D) for 64-bit elements. With an SVL of 512 bits, each ZA.S tile has dimensions of $16 \times 16$, while each ZA.D tile has dimensions of $8 \times 8$.
Each ZA tile can be accessed as a whole or in slices, with slices composed of horizontally or vertically arranged vectors.

\begin{table}[t]
    \centering
    \caption{ZA tiles for different element types.}
    \label{tab:za_tile}
    \small
    \begin{tabular}{l l l}
    \toprule
    Element types & Number of tiles & Tile names \\
    \midrule
    8-bit    & 1     & ZA0.B \\
    16-bit   & 2     & ZA0.H-ZA1.H \\
    32-bit   & 4     & ZA0.S-ZA3.S \\
    64-bit   & 8     & ZA0.D-ZA7.D \\
    128-bit  & 16    & ZA0.Q-ZA15.Q \\
    \bottomrule
    \end{tabular}
    \vspace{-4mm}
\end{table}

The fundamental primitives of SME are outer product and accumulate instructions, which enable the construction of matrix multiplication in software. These instructions take two Z registers as inputs and accumulate results into a ZA tile, supporting both integer and floating-point formats across multiple precisions. For example, in the FP32 case, the floating-point outer product and accumulate instruction (FMOPA) computes the outer product of two $\tfrac{SVL}{32}$-element vectors and accumulates the result into a ZA tile with $\tfrac{SVL}{32} \times \tfrac{SVL}{32}$, performing $2 \times (\tfrac{SVL}{32})^2$ floating-point operations. In addition, SME outer product instructions allow low-precision inputs with high-precision accumulation, including FP16/BFloat16 $\rightarrow$  FP32, INT8 $\rightarrow$ INT32, and INT16 $\rightarrow$  INT32/INT64. This feature enhances computational intensity and reduces memory bandwidth pressure, enabling efficient mixed-precision GEMM with lower-precision inputs and higher-precision outputs~\cite{mixed-precision, QuantizedNN,ladder}.

\subsection{GEMM Algorithm and its Mapping to SME}
\begin{figure}[t]
    \centering
    \begin{subfigure}{0.49\textwidth} 
        \centering
    \includegraphics[width=0.9\linewidth]{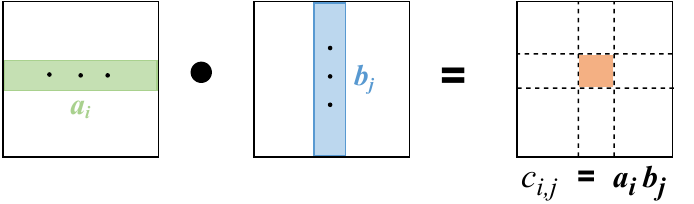}
        \caption{Inner product}
        \label{fig:inner}
    \end{subfigure}
    \begin{subfigure}{0.49\textwidth}
        \centering
    \includegraphics[width=0.9\linewidth]{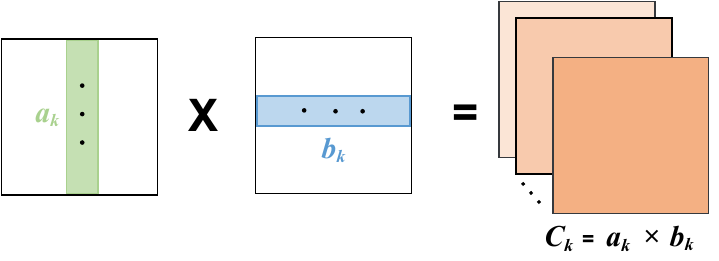}
        \caption{Outer product}
        \label{fig:outer}
    \end{subfigure}
    \caption{The inner product and outer product of GEMM.} 
    \vspace{-4mm}
    \label{fig:gemm_product}
\end{figure}

GEMM algorithms can be categorized into inner product and outer product approaches, depending on the loop traversal order. 
In Fig.~\ref{fig:inner}, the inner product method computes each element $c_{i,j}$ of matrix $C$ by taking the dot product between a row vector $\mathbf{a}_{i}$ and a column vector $\mathbf{b}_{j}$, i.e., $c_{i,j} = \mathbf{a}_{i} \cdot \mathbf{b}_{j} =
 \sum_{k=1}^{K} a_{i,k} b_{k,j}$. 
In contrast, the outer product method traverses along the $K$ dimension described in Fig.~\ref{fig:outer}. At each step $k$, the $k$-th column vector $\mathbf{a}_{k}$ and the $k$-th row vector $\mathbf{b}_{k}$ form the outer product, generating a rank-1 matrix $\mathbf{C}_{k} = \mathbf{a}_{k} \times \mathbf{b}_{k}$, which is then accumulated into $\mathbf{C} = \sum_{k=1}^{K} \mathbf{C}_{k}$.

The outer product offers a higher compute-to-memory ratio than the inner product. Moreover, its pattern naturally aligns with the FMOPA instruction. Firstly, a column vector $\mathbf{a}_{k}$ and a row vector $\mathbf{b}_{k}$ are loaded into registers. Secondly, the FMOPA instruction computes their outer product and accumulates the results into the ZA storage. Finally, the ZA storage results are written back to matrix $\mathbf{C}$. This synergy between hardware and algorithms enables SME to fully exploit its hardware potential for GEMM computations.

\subsection{Motivation: Limitations of Existing Implementations}
Despite the clear hardware advantages, our analysis reveals that current SME-enabled GEMM libraries fail to fully leverage SME’s potential, particularly for large matrices. We identify three critical shortcomings that motivate our work.

\subsubsection{Ineffective cache hierarchy utilization} 
Existing open-source libraries~\cite{libxsmm, openblas, kleidiai} rely on straightforward three-level loops and do not employ multi-level blocking strategies. Their fixed tile sizes often fail to fit the shared L2 cache effectively. Therefore, we transform the original three-level loops into a six-level blocked loop structure. Tiling parameters are guided by an analytical model that considers the L2 cache size, TLB size, and memory bandwidth to maximize cache locality after the packing stage.

\subsubsection{Suboptimal data packing method} 
LIBXSMM and OpenBLAS pack only one input matrix to reduce packing overhead while maintaining the layout required by outer product instructions. However, if the unpacked matrix exceeds the shared L2 cache capacity, performance degrades significantly.
To overcome this drawback, we pack both input matrices to decrease the stride between consecutive accesses and improve spatial locality. Consecutive accesses enable full utilization of the shared L2 cache banks, thereby maximizing parallelism and hardware prefetching. In addition to on-the-fly fast transposition, we further apply the first-round online packing strategy to amortize memory overhead.

\subsubsection{Underutilization of multi-vector load instructions} 
SME2 supports multi-vector data processing instructions. As shown in Fig.~\ref{fig:bandwidth}, when each load transfers less than 8 MB, employing four Z registers per group (256 bytes) can achieve up to 900 GB/s, whereas a single Z register (64 bytes) delivers only 230 GB/s. However, existing GEMM packing methods and micro-kernels utilize only one or two Z registers per load, which constrains both memory bandwidth and computational throughput. Therefore, we propose restructuring the micro-kernel to consistently load data using four Z-register groups.

These limitations collectively result in suboptimal performance for large-scale GEMMs. This gap motivates the design of \SystemName, which systematically addresses these issues through cache-aware partitioning, efficient dual-matrix packing, and micro-kernels designed to exploit high-bandwidth multi-vector loads.

\section{Microbenchmarks} 
\label{sec:microbenchmark}

To guide the design of \SystemName, we conduct a series of microbenchmarks on the Apple M4 Pro chip to characterize the performance properties of its SME units. These experiments reveal critical architectural features that directly inform our optimization strategies.

\begin{figure}[t]
    \centering
    \begin{subfigure}{0.49\textwidth} 
        \centering
    \includegraphics[width=1\linewidth]{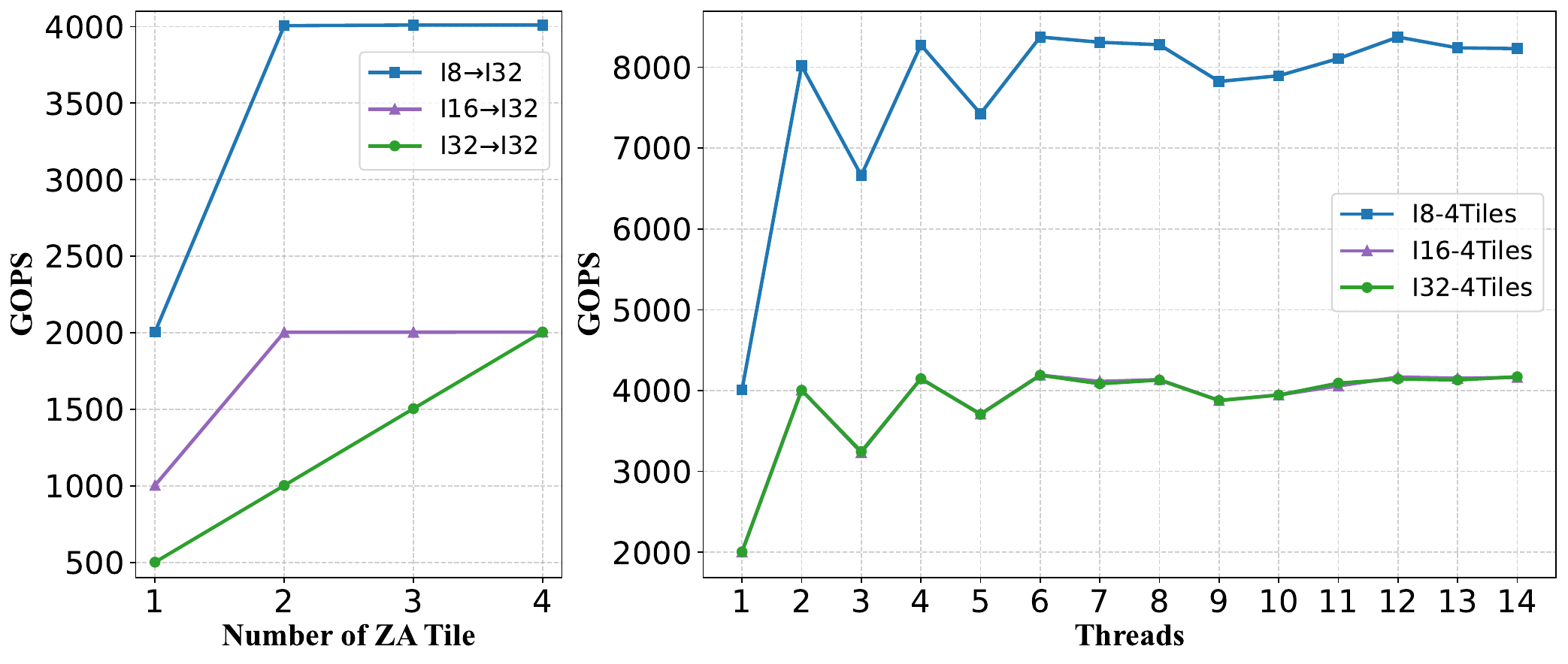}
        \caption{Integer outer product and accumulate instructions.}
        \label{fig:zareuse-int}
    \end{subfigure}
    
    \vspace{2mm}
    
    \begin{subfigure}{0.49\textwidth}
    \centering
    \includegraphics[width=1\linewidth]{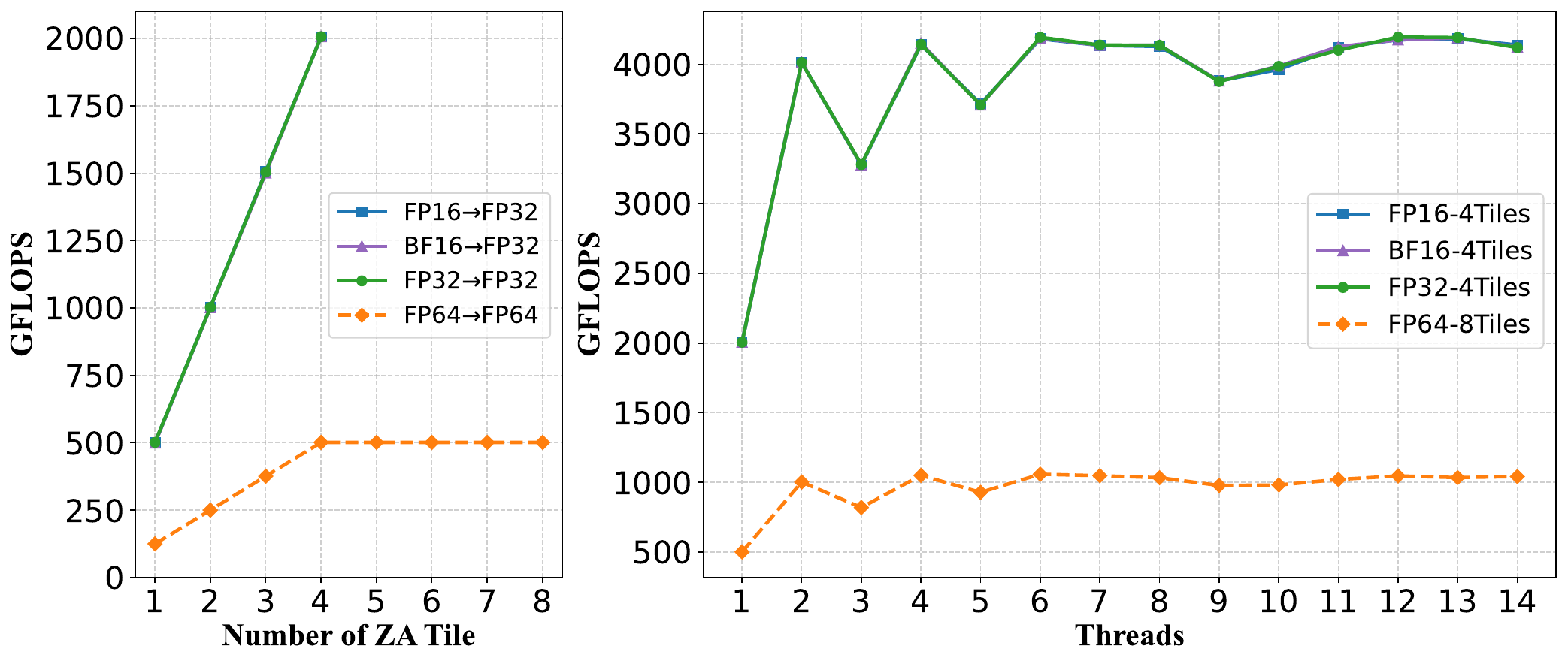}
        \caption{Floating-point outer product and accumulate instructions.}
        \label{fig:zareuse-float}
    \end{subfigure}
    \caption{Peak performance of outer product and accumulate instructions under varying numbers of ZA tiles and thread counts.} 
    \vspace{-4mm}
    \label{fig:zareuse}
\end{figure}

\subsection{Charaterizing Computational Throughput} 
We first measure the peak floating-point throughput of the SME unit by varying the number of active ZA tiles for outer product operations. For 32-bit elements (FP32 and INT32), the ZA storage can be partitioned into four tiles (ZA0.S–ZA3.S), whereas for 64-bit elements it can be partitioned into eight tiles. We evaluate common outer product accumulation operations: INT8/INT16/INT32 $\rightarrow$ INT32, FP16/BF16/FP32 $\rightarrow$ FP32, and FP64 $\rightarrow$ FP64.

As shown in Fig.~\ref{fig:zareuse}, utilizing four tiles ensures reaching the maximum throughput. This clearly demonstrates that maximizing performance requires designing the micro-kernel to utilize all ZA tiles concurrently.
The INT8$\rightarrow$INT32 outer product instruction achieves 4010 GOPS with four ZA tiles in a single thread, whereas all other precisions (INT16, INT32, FP16, BF16, and FP32) peak at 2006 GOPS or GFLOPS. 
However, FP64 performance is limited to 501 GFLOPS, representing only one-quarter of the throughput of other non-INT8 precisions (Fig.~\ref{fig:zareuse-float}).

We further evaluate multi-threaded scaling using Apple's Dispatch framework. Unlike Linux, macOS does not allow explicit thread-core binding. 
With two threads, the INT8 outer product performance doubles to 8020 GOPS, as shown in Fig.~\ref{fig:zareuse-int}. As the thread count increases, the scheduler attempts to distribute threads across clusters for load balancing, and throughput peaks at 8375 GOPS. 
Other precisions show the same trend, reaching 4012 GOPS with two threads and up to 4195 GOPS at peak. 
This scaling behavior suggests that the M4 Pro contains two high-throughput SME units associated with the P-core clusters and one lower-throughput unit with the E-core cluster, with cores within a cluster sharing a single SME unit. Contention occurs when multiple threads access the same SME unit simultaneously, capping the peak throughput.

\begin{figure}[t]
    \centerline{\includegraphics[width=1\linewidth]{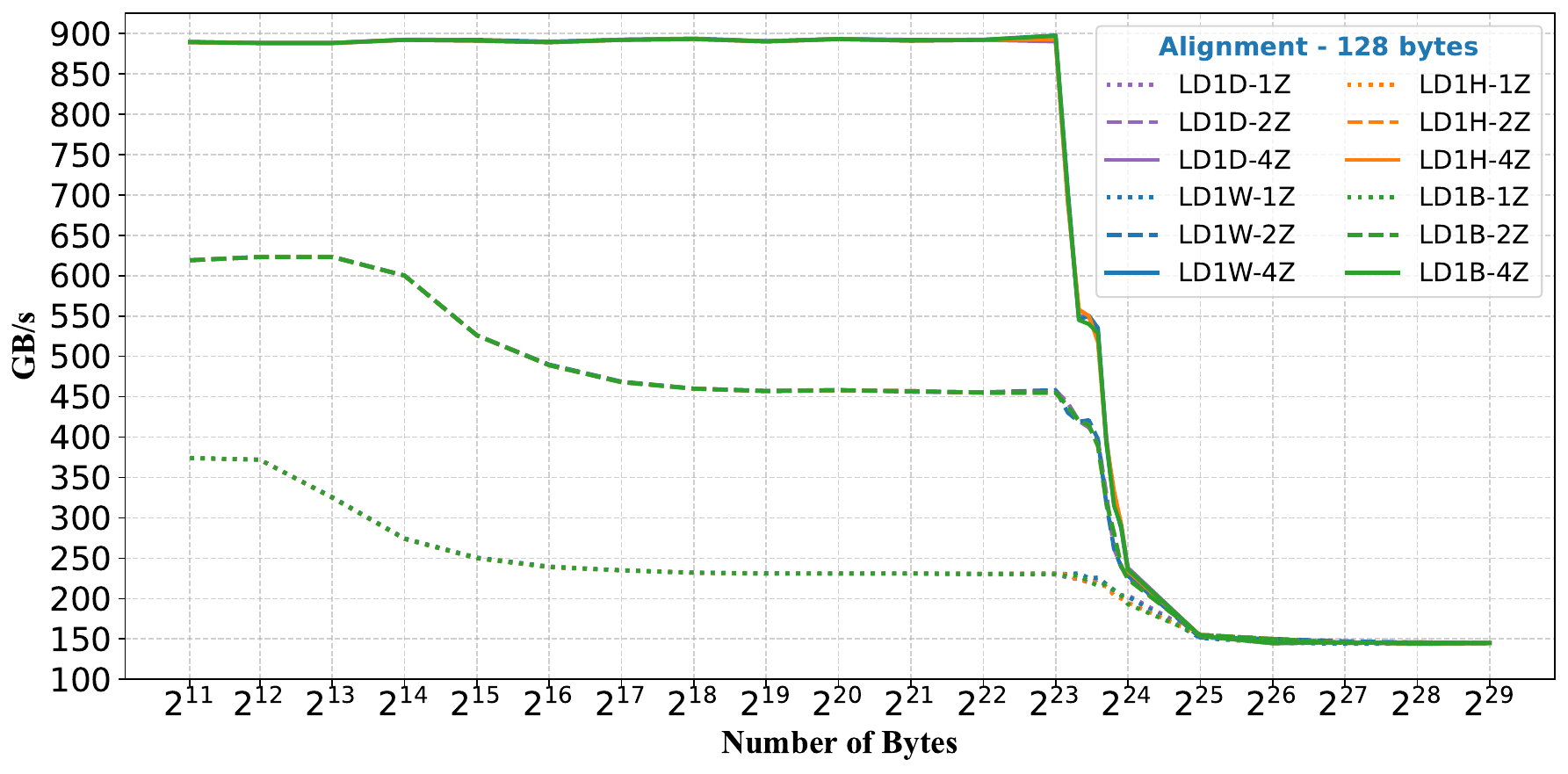}
    }
    \caption{Memory bandwidth for loading data into the ZA storage using LD1D, LD1W, LD1H, and LD1B instructions, which load 64-, 32-, 16-, and 8-bit data elements, respectively, with 1, 2, or 4 Z registers. The horizontal axis represents the total bytes loaded.}
    \label{fig:bandwidth}
    \vspace{-4mm}
\end{figure}

\subsection{Charaterizing Memory Bandwidth} 

A key bottleneck for GEMM performance is often memory bandwidth. We benchmark the effective bandwidth achievable by the SME unit for different vector load methods, measuring the rate of loading data from memory into the ZA storage.

The results, presented in Fig.~\ref{fig:bandwidth}, reveal a critical insight: bandwidth is determined by the number of Z registers loaded per operation. Loading a single Z register (64 bytes) achieves 375 GB/s, while loading groups of two (128 bytes) and four (256 bytes) registers achieve 625 GB/s and 900 GB/s, respectively. 
As bus pressure increases, the bandwidth for loading one and two Z registers per group gradually decreases to 230 GB/s and 460 GB/s, respectively.
This underscores the necessity of using four-Z-register load instructions to saturate the memory bandwidth.

However, this peak bandwidth is contingent on data locality. When the total loaded data exceeds 8 MB, bandwidth drops precipitously to approximately 230 GB/s, with a further decline to 145 GB/s beyond 16 MB. This indicates that the SME unit's effective cacheable working set is limited to 8 MB for optimal bandwidth. In contrast, write-back bandwidth to memory peaks at 230 GB/s and is unaffected by multi-vector stores. These findings directly inform our partitioning and packing strategy: submatrices must fit within an 8 MB footprint, and data must be loaded using four-Z-register groups.

\begin{figure}[t]
\centerline{\includegraphics[width=1\linewidth]{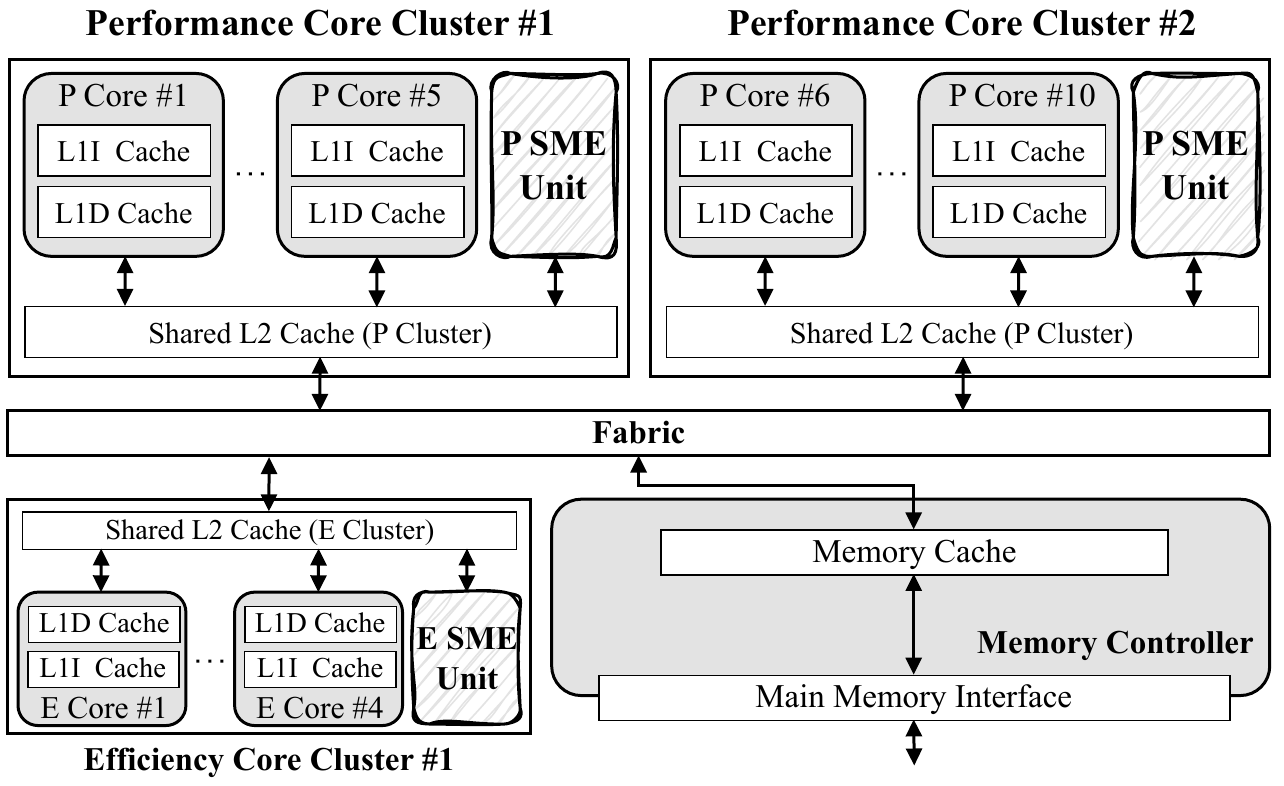}
}
\caption{A high-level view of the Apple M4 Pro architecture.}
\label{fig:SME-view} 
\vspace{-4mm}
\end{figure}

\subsection{Architectural Implications and Optimization Guidelines}

Based on our microbenchmarks and hardware specifications, we infer the high-level architecture of the Apple M4 Pro, depicted in Fig.~\ref{fig:SME-view}. Each cluster of cores (performance and efficiency) integrates a shared SME unit, private L1 caches, and a shared L2 cache. All clusters are connected to the memory subsystem via an on-chip fabric.  We further observe that a memory ordering relationship exists between the CPU core and the SME unit. Simultaneous access to the same memory by both can cause severe pipeline stalls. Therefore, data buffers for SME operations should be allocated in dedicated heap memory, avoiding placement on the stack or in shared regions with core-resident data.

To summarize, our microbenchmarks yield three fundamental principles for optimizing GEMM on this architecture: (1) \textbf{Maximize Tile Usage}: Micro-kernels utilize all available ZA tiles to ensure peak computational throughput is achieved. (2) \textbf{Optimize Load Granularity}: Consistently use four-Z-register load instructions to maximize memory bandwidth. (3) \textbf{Respect Cache Limits}: Partition data to ensure submatrices remain within an 8 MB working set to maintain high bandwidth. These principles form the foundation of \SystemName's design, which we detail in the following sections.

\begin{figure}[!t]
\centerline{\includegraphics[width=1\linewidth]{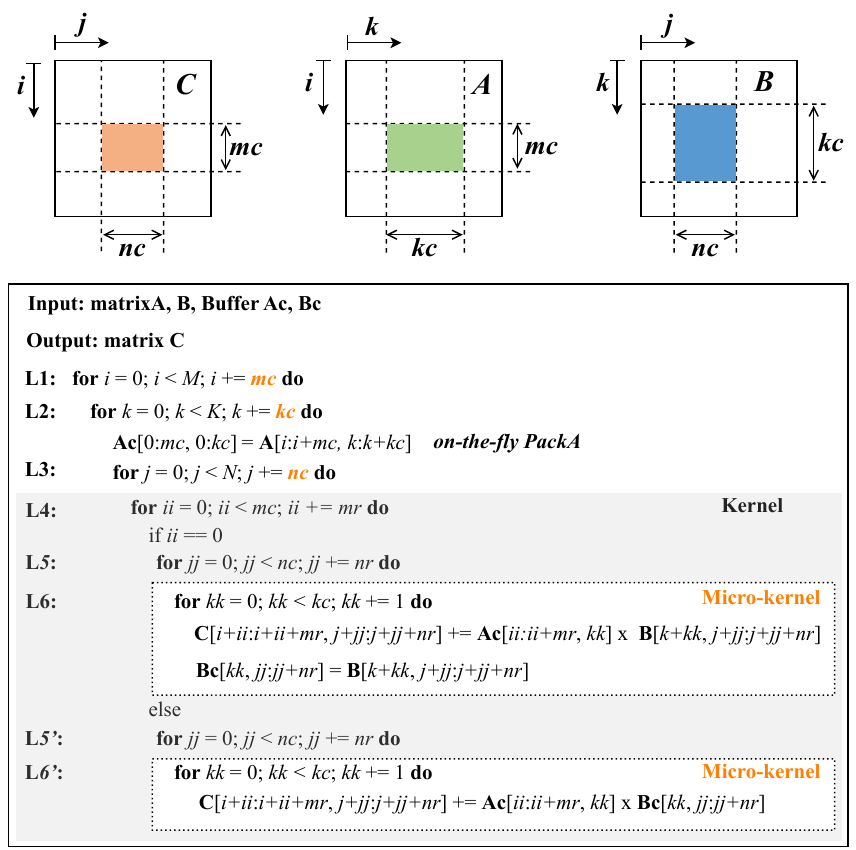}
}
\caption{Algorithm design of \SystemName .}
\label{fig:algorithm} 
\vspace{-4mm}
\end{figure}

\section{Optimizing General Matrix Multiplication} \label{sec:design}
Building on our architectural insights, we now present the design of \SystemName for double-precision and single-precision GEMM. We use the row-major FP32 GEMM operation $\mathbf{C} = \alpha \mathbf{A} \cdot \mathbf{B} + \beta \mathbf{C}$ on a platform with SVL=512 bits as a running example. Our methodology generalizes to other platforms, precisions, and storage orders.

\subsection{Design Overview}
\SystemName is built upon the classical \textit{Goto} algorithm but introduces specialized adaptations for SME. As shown in Fig.~\ref{fig:algorithm}, the computation is structured around a multi-level blocking strategy designed to exploit the memory hierarchy.

\SystemName partitions matrices $\mathbf{A}$, $\mathbf{B}$, and $\mathbf{C}$ into small submatrices across loops L1-L3 in Fig.~\ref{fig:algorithm}. Memory access efficiency and cache utilization are enhanced through two packing operations that rearrange the $mc \times kc$ and $kc \times nc$ submatrices of $\mathbf{A}$ and $\mathbf{B}$ into contiguous buffers $Ac$ and $Bc$. The tiling parameters $mc$, $nc$, and $kc$ are guided by an analytical model to maximize cache locality. We leverage SME's on-the-fly transposition to efficiently pack $\mathbf{A}$ and employ a first-round online packing method for $\mathbf{B}$, mitigating the memory overhead of packing.
Loops L4-L6 perform the core matrix multiplication using a three-nested loop. In L4, $Ac$ is divided into a series of $mr \times kc$ row panels $Ar$, and in L5, $Bc$ is divided into $kc \times nr$ column panels $Br$, enabling the efficient use of vector load and outer product instructions. Loop L6/L6' implement the GEMM micro-kernel, the well-known core in BLAS libraries. The micro-kernel is based on an outer product approach, where the FMOPA instruction computes the outer product between the column-major order $mr \times 1$ slice of $Ar$ and the row-major order $1 \times nr$ slice of $Br$, accumulating the results into the $mr \times nr$ sub-block $Cr$ stored in the ZA storage. The parameters $mr$ and $nr$ are critical for performance, as they determine both the ZA tiles and the Z registers used per load when updating the sub-block $Cr$. 
We parallelize the $m$ and $n$ dimensions of loops L1 and L3 depicted in Fig.~\ref{fig:algorithm} to utilize multiple SME units for accelerating GEMM. Since the $K$ dimension is the reduction dimension in matrix multiplication and introduces write-after-write dependencies, loop L2 is not parallelized. Both the packing routines and the micro-kernels are implemented in native SME assembly.

\subsection{Cache-Aware Partitioning and Packing}
Tiling parameters are critical for effective utilization of the cache hierarchy~\cite{partition}. Since SME only shares the L2 cache with CPU cores, we map blocks that traditionally reside in L2/L3 caches to the shared L2 cache, without reserving L1 cache space. The buffers $Ac$ and $Bc$ must be aligned to cacheline boundaries to fully exploit SME performance.
Within loop L2, $\mathbf{A}$ is partitioned into $mc \times kc$ submatrices and rearranged into the buffer $Ac$; in loop L3, $\mathbf{B}$ and $\mathbf{C}$ are divided into $kc \times nc$ and $mc \times nc$ submatrices, respectively. We expect the L2 cache to hold these blocks. 
In addition, the current CPU adopts an LRU cache replacement policy, which evicts previously used data to accommodate newly generated data.
It is necessary to reserve space for the buffer $Bc$ and the submatrix of $\mathbf{C}$, ensuring that $Bc$ is not evicted from the L2 cache during the next iteration of loop L3, thus imposing equation~\eqref{eq:partition}. 
In our platform, we constrain the memory footprint of these blocks under 8 MB, as microbenchmarks show.

Our micro-kernel employs an outer product approach. To enable consecutive memory access, submatrices are packed: $\mathbf{A}$ into column-major $mr \times kc$ panels ($Ar$ in $Ac$) and $\mathbf{B}$ into row-major $kc \times nr$ panels ($Br$ in $Bc$). The tile sizes $mr = 16$ and $nr = 64$ are chosen according to the micro-kernel design (Section~\ref{sec:micro-kerneldeisgn}). Address translation is handled by the core load/store unit, with both the core L1D TLB and the shared L2 TLB covering SME unit accesses. The large L2 TLB can be ignored. Under constraint~\eqref{eq:tlb}, TLB misses are preferred not to occur. Here, $2 \times Tb$ covers the current and next loop iterations. $Tc$ is not doubled because each row of size $mr \times nr$ sub-block may reuse the same TLB entry across iterations.
Once $mr$ and $nr$ are fixed, we determine $kc$ via the TLB-aware equation~\eqref{eq:tlb}. 
We then use Lagrange multipliers to maximize the L2 compute-to-memory ratio (CMR) defined in equation~\eqref{eq:cmr}, subject to the L2-size constraint in equation~\eqref{eq:partition}. 
This yields $mc$, $nc$, and $kc$ that balance cache residency and computational intensity.

\begin{equation} 
\begin{aligned} 
& mc \times kc + kc \times nc + mc \times nc \\ &+ kc \times nc  + mc \times nc < \frac{L2 \ Cache}{dtype\_size} 
\end{aligned} 
\label{eq:partition} 
\end{equation}

\begin{equation}
\left\{
\begin{aligned}
& Ta + 2 \times Tb + Tc < T\_entry\_total, \\
& Ta = \Bigl\lceil \frac{mr \times  kc \times dtype\_size}{page\_size} \Bigr\rceil + 1, \\
& Tb = \Bigl\lceil \frac{nr \times kc \times dtype\_size}{page\_size} \Bigr\rceil + 1, \\
& Tc \le mr \quad (  ldc \text{ is large})
\end{aligned}
\right.
\label{eq:tlb}
\end{equation}

\begin{equation} 
\begin{aligned} 
& CMR = \frac{2 \times mc \times nc \times kc}{mc \times kc + kc \times nc + 2 \times mc \times nc} 
\end{aligned} 
\label{eq:cmr} 
\end{equation}

\begin{figure}[t]
\centerline{\includegraphics[width=1\linewidth]{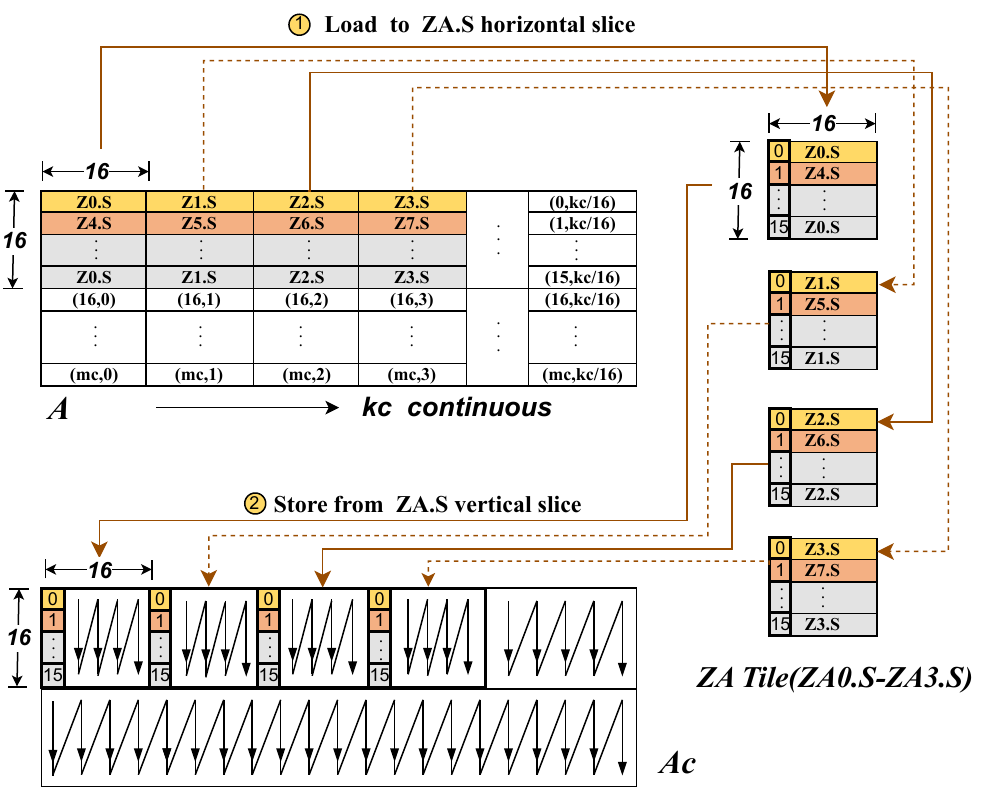}
}
\caption{The on-the-fly transposition of the submatrix of $A$ is realized by the ZA tile. The row vectors are loaded into the horizontal slices and then written via the vertical slices.}
\label{fig:packa-fp32} 
\vspace{-4mm}
\end{figure}

To minimize the overhead of data packing, we introduce two optimized strategies.

\cparagraph{On-the-fly Transposition for Matrix A} 
With ZA tiles supporting both horizontal and vertical slicing, we perform on-the-fly transposition of submatrices from $\mathbf{A}$ in Fig.~\ref{fig:packa-fp32}. 
For FP32 elements, the ZA storage can be viewed as four tiles, with each tile holding $16 \times 16$ values. 
The transformation proceeds in panels of $16 \times kc$ along the $M$ dimension.
Within each panel, the data are processed in $16 \times 64$ sub-panels along $K$.
For each row of a $16 \times 64$ sub-panel, four Z registers are loaded
in one instruction and written horizontally into the four ZA tiles.
After all $16$ rows are loaded, the ZA tiles are fully populated.
The transposition is realized by vertically slicing the ZA tiles to extract
four column-major $16 \times 16$ chunks, which are written to the buffer
$Ac$.
Repeating this process over all sub-panels yields a $16 \times kc$ column-major
panel, and iterating over the $M$ dimension produces the final packed layout.
Predicate registers are used to mask inactive lanes when handling boundary cases.
This approach eliminates inefficient gather operations and improves memory access efficiency.

\cparagraph{First-round Online Packing for Matrix B} Matrix $\mathbf{B}$ is already in row-major order storage, so no transposition is needed. Since the computation instructions are independent from the load/store channels, we overlap the load and store operations of the packing stage with the FMOPA instructions during each iteration of loop L5 in Fig.~\ref{fig:algorithm}, reducing memory access latency. In this iteration, while using FMOPA to update the $mr \times nr$ sub-block of $\mathbf{C}$, the corresponding $kc \times nr$ panel of $B$ is sequentially written into the linear buffer $Bc$. The entire $kc \times nc$ buffer $Bc$ is constructed by the end of the L5 loop. To maximize memory bandwidth, we load four Z registers per group (64 FP32 elements) during both packing and computation. In subsequent iterations of L4, updates to the $mr \rightarrow mc$ blocks of $\mathbf{C}$ reuse this buffer, reducing the packing latency and preserving continuous memory access.


\begin{figure}[t]
\centerline{\includegraphics[width=1\linewidth]{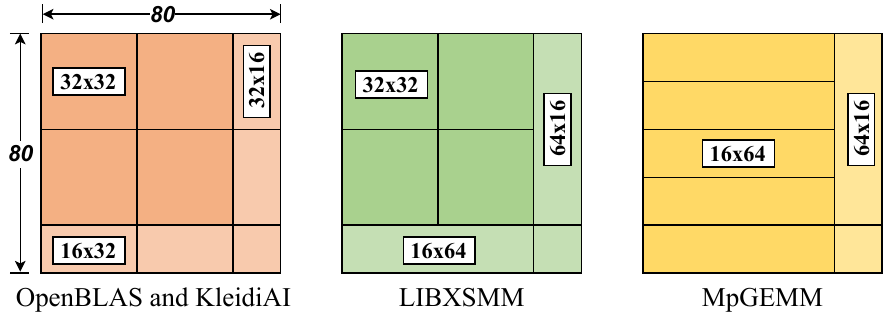}
}
\caption{The shape of micro-kernel. (left) 32x32 micro-kernel used in OpenBLAS and KleidiAI; (middle) 32x32 main micro-kernel and 16x64 and 64x16 edge micro-kernels in LIBXSMM; (right) 16x64 main micro-kernel and 64x16 edge micro-kernel in \SystemName .}
\label{fig:fp32-shape} 
\vspace{-4mm}
\end{figure}

\subsection{High-Performance Micro-Kernel Design}
\label{sec:micro-kerneldeisgn}
We design main and edge micro-kernels to utilize all ZA tiles and four-Z-register load instructions. In FP32 arithmetic, ZA storage can be viewed as four ZA tiles, each of size $\tfrac{SVL}{32} \times \tfrac{SVL}{32}$. 
Microbenchmarks demonstrate that SME throughput scales with the number of active ZA tiles. With an SVL of 512 bits, utilizing all four ZA tiles enables three valid micro-kernel configurations: $32 \times 32$, $16 \times 64$, and $64 \times 16$.
Previous micro-kernels adopt a $32 \times 32$ layout shown in Fig.~\ref{fig:fp32-shape}.
LIBXSMM improves edge handling with $16 \times 64$ and $64 \times 16$ micro-kernels, enabling parallel updates across all four ZA tiles, while OpenBLAS and KleidiAI use only two tiles at the boundaries.

However, bandwidth limits stem from suboptimal multi-vector load utilization.
In detail, the sub-blocks $Cr$ are loaded into ZA tiles for outer product accumulation and then written back, resulting in repeated loads and stores along the reduction dimension due to multi-level blocking.
For $nr = 32$, only two Z registers can be loaded per group and transferred to ZA tiles via Mov instructions, whereas $nr = 64$ allows four Z vector registers to be loaded simultaneously. For large-scale GEMMs, repeated accesses to $Cr$ make their bandwidth efficiency the performance bottleneck.
The bandwidth achieves only 460 GB/s for two-register group loads, compared to 900 GB/s for four-register group loads.
Therefore, we adopt a $ mr \times nr = 16 \times 64$ main micro-kernel (and $64 \times 16$ edge micro-kernel) that fully utilizes all ZA tiles and four-Z-register loads.
In the column-major layout, the roles are reversed, with a $mr \times nr = 64 \times 16$ main micro-kernel and a $16 \times 64$ edge micro-kernel. This design principle generalizes to platforms where the SVL differs from 512 bits.

\begin{algorithm}[t] \scriptsize
\caption{Main Micro-Kernel}
\label{alg:fp32main_kernel}
$(ZA0.S-ZA3.S) \leftarrow \textbf{Cr}(0:16,0:64)$ \tcp{4-way loading}  

\For{$kk=0$ $\rightarrow$ $kc$ \textbf{step}=16}{
    \tcp{step = 0}
    $(Z0.S-Z3.S) \leftarrow \textbf{Ar}(0:16, kk:kk+4)$  
    \\
    \tcp{Pipelined outer product updates (interleave load and FMOPA)}
    
    $(Z4.S-Z19.S) \leftarrow \textbf{Br}(kk:kk+4, 0:63)$ 
    \\
    \tcp{loading B in batches of 4 Z registers}  
    
    \For{$i=0$ to 3}{
        $(ZA0.S-ZA3.S) \leftarrow \text{FMOPA} \ Z_i.S, (Z_{4+4i}.S \sim Z_{7+4i}.S)$  \\
        \tcp{while computing, next batch of A/B can start loading}
    }
    \tcp{step = 4 to 12}
}
$\textbf{Cr}(0:16,0:64) \leftarrow (ZA0.S-ZA3.S)$  
\end{algorithm}
We first load the sub-blocks $Cr$ into Z registers in a four-Z-register way and then move them into the ZA tiles to maximize bandwidth.
The data of $Ar$ and $Br$ are also loaded in batches using four Z registers. As illustrated in Algorithm~\ref{alg:fp32main_kernel}, in step 0, four columns of $Ar$ are loaded into Z0.S-Z3.S, and the first row of $Br$ into Z4.S-Z7.S, performing four outer product operations that accumulate into four ZA tiles. Subsequently, the following rows of $Br$ (Z8.S-Z11.S, Z12.S-Z15.S, Z16.S-Z19.S) are loaded in turn and multiplied with Z1.S-Z3.S to update the results. 
Due to the latency associated with accessing data from the shared L2 cache, issuing dependent FMOPA instructions immediately after corresponding load operations reduces pipeline efficiency. To address this issue, we employ loop unrolling and software pipelining techniques: on the one hand, we launch subsequent iteration load instructions early, allowing FMOPA operations from the previous iteration to overlap with the load latency; on the other hand, we utilize more Z registers to buffer data, thereby alleviating scheduler pressure.
At the end of the loop, the results in the ZA tiles are written back to the sub-blocks $Cr$.



\section{Mixed-Precision GEMM Kernel Design}\label{sec:mpdesign}

\subsection{Design Principals}
The computational demands of deep learning have driven the adoption of lower-precision formats like FP16 and INT8 to reduce memory footprint and increase throughput. ARM SME natively supports mixed-precision operations through instructions that take lower-precision inputs and accumulate into higher-precision outputs (e.g., FP16$\rightarrow$FP32, INT8$\rightarrow$INT32). 
This section details how we extend \SystemName's optimization strategies to leverage these capabilities efficiently.
Here, we take mixed-precision GEMM with FP16 inputs and FP32 outputs as an example.
The fundamental workflow for mixed-precision GEMM remains consistent with the FP32 case, as outlined in Fig.~\ref{fig:algorithm}. 
The primary challenge lies in adapting the data packing and micro-kernel to handle the different data layouts and computational patterns of mixed-precision operations. 

\begin{figure}[t]
\centerline{\includegraphics[width=1\linewidth]{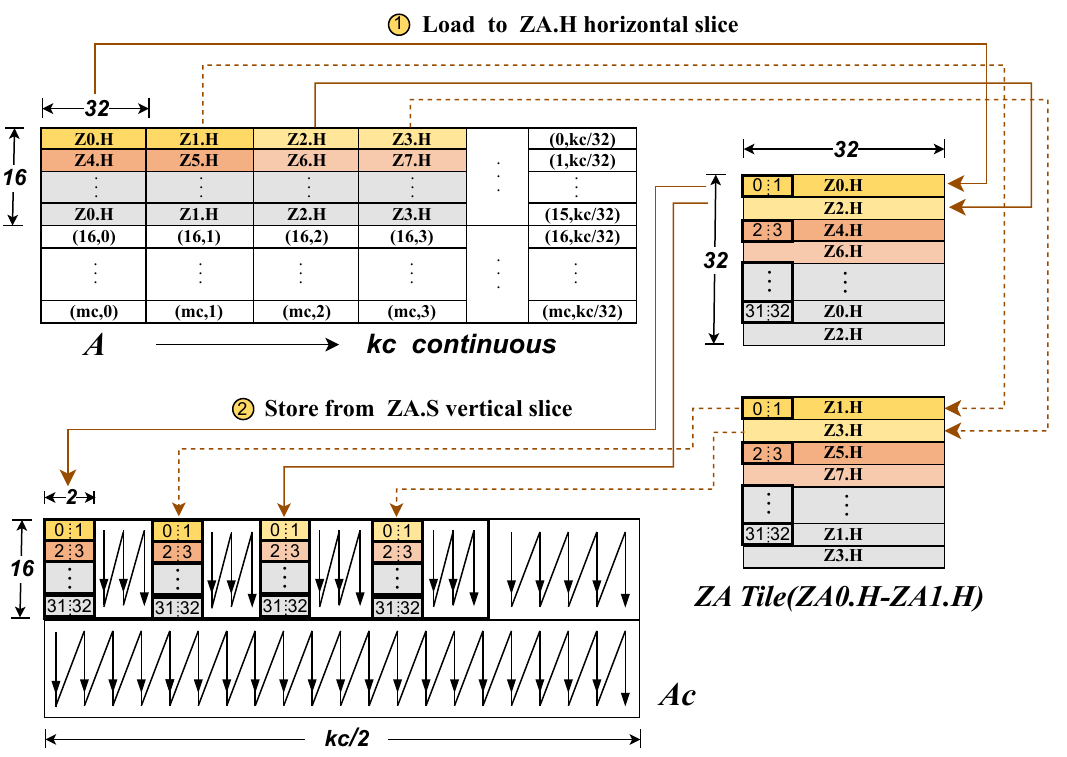}
}
\caption{Transposed packing of an FP16 submatrix of $A$.}
\label{fig:packa-fp16} 
\vspace{-4mm}
\end{figure}

\begin{figure}[t]
\centerline{\includegraphics[width=1\linewidth]{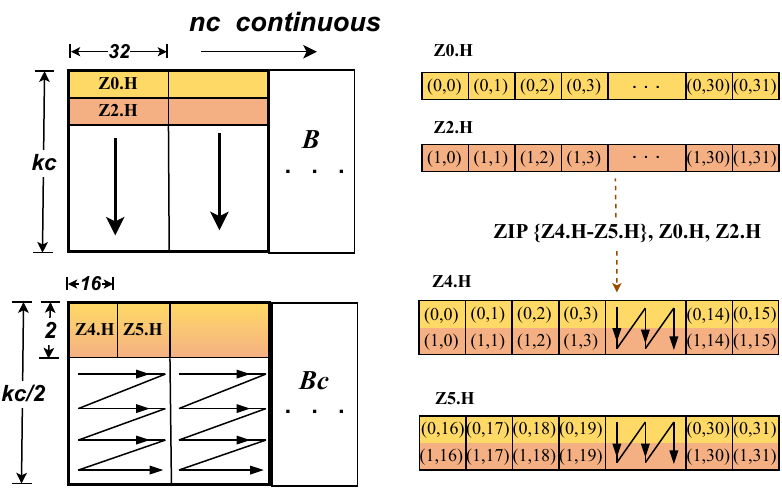}
}
\vspace{0mm}
\caption{Vertical interleave packing of an FP16 submatrix of $B$.}
\label{fig:packb-fp16} 
\vspace{-2mm}
\end{figure}

\subsection{Packing Strategies for Lower Precision}

For FP16 data type, we still apply the on-the-fly transposition for matrix $A$. We treat two consecutive FP16 elements in the horizontal dimension as a FP32 value. When the element type is FP16, the ZA storage can be viewed as comprising two ZA tiles (ZA0.H and ZA1.H). Both tiles, along with ZA0.S-ZA3.S, feature interleaved horizontal tile slices to improve hardware efficiency in accessing ZA tiles and tile slices. 
In Fig.~\ref{fig:packa-fp16}, we load four $Z$ registers (Z0.H-Z3.H, totaling 128 FP16 elements) at a time and interleave them across the horizontal slices of ZA0.H and ZA1.H. 
When accessing the tile via FP32 vertical slices, two consecutive FP16 elements are treated as a single FP32 value. This operation physically interleaves accesses to ZA0.H and ZA1.H. Subsequently, the four vertical slices are written sequentially to the linear buffer $Ac$.
In this process, the original $16 \times 128$ row-major FP16 elements are rearranged into a column-major layout. 
Iterating along the $kc$ and $mc$ dimensions completes the transposed packing of the entire $mc \times kc$ submatrix.

During the packing of matrix $B$, the ZIP instruction provided by SME is employed for efficient data interleaving. As illustrated in Fig.~\ref{fig:packb-fp16}, the ZIP instruction interleaves source vector registers Z0.H and Z2.H and writes the results into a pair of destination registers Z4.H and Z5.H. Specifically, the first 16 FP16 elements of Z0.H and Z2.H are interleaved into Z4.H, while the remaining 16 elements are interleaved into Z5.H. 
Adjacent rows of matrix $B$ are vertically interleaved, combining every two FP16 elements into a logical FP32 value.
In practice, vectors from the $kk$-th and $(kk+1)$-th rows are loaded, interleaved into the destination registers, and then mapped into the buffer $Bc$. We still need to use predicate registers to mask the tail blocks. 
We iterate along the $kc$ and $nc$ dimensions to rearrange the data into $\tfrac{nc}{32}$ row-major $kc \times 32$ sub-blocks. 
This layout ensures contiguous, aligned access patterns during the micro-kernel computation.


\begin{algorithm}[t] \scriptsize
\caption{Mixed-Precision Micro-Kernel}
\label{alg:fp16main_kernel}
$(ZA0.S-ZA3.S) \leftarrow \textbf{Cr}(0:16,0:64)$ \tcp{4-way loading}  
\For{$kk=0$ $\rightarrow$ $kc/2$ \textbf{step}=16}{
    \tcp{step = 0}
    $(Z0.H-Z3.H) \leftarrow \textbf{Ac}(0:16, kk:kk+4)$  
    \\
    \tcp{Pipelined outer product updates (interleave load and FMOPA)}
    $(Z4.H-Z19.H) \leftarrow \textbf{Bc}(kk:kk+4, 0:63)$ 
    \\
    \tcp{loading B in two halves (0:31,32:63)}
    \For{$i=0$ to 3}{
        $(ZA0.S-ZA1.S) \leftarrow \text{FMOPA}\ Z_i.H, (\text{corresponding B batch})$  
        \\
        $(ZA2.S-ZA3.S) \leftarrow \text{FMOPA}\ Z_i.H, (\text{corresponding B batch})$  
        \\
        \tcp{while computing, next batch of A/B can start loading}
    }
    \tcp{step = 4 to 12}
}
$\textbf{Cr}(0:16,0:64) \leftarrow (ZA0.S-ZA3.S)$  
\end{algorithm}

\subsection{Mixed-Precision Micro-Kernel}
The mixed-precision micro-kernel is designed to exploit the four-Z-register load and FP16 FMOPA instruction. This FMOPA instruction can be viewed as performing a multiplication between a 16×2 FP16 matrix and a 2×16 FP16 matrix, producing a 16×16 FP32 accumulated result. Following this computation pattern, the micro-kernel uses a 16×64 tile size ($mr$ = 16, $nr$ = 64)  to collaboratively update the output with four 16×16 FP32 ZA tiles (ZA0.S-ZA3.S). When loading a FP32 $16 \times 64$ sub-block $Cr$, \SystemName  still loads four Z registers per group. Loads are masked using predicate registers when fewer than 64 values are required.

The outer product computation along a loop kk with a step size of 16 described in Algorithm~\ref{alg:fp16main_kernel}. In step 0, we load the four columns of buffer $Ac$ into Z0.H-Z3.H, with each column containing 32 FP16 values. 
Buffer $Bc$ is organized into a series of $kc \times 32$ panels.  
To accommodate a micro-kernel size of $16 \times 64$ and the four-column-per-load layout of $Ac$, we sequentially load the four rows of two adjacent $kc \times 32$ panels from buffer $Bc$.  
These data are loaded into 16 Z registers (e.g., Z4.H-Z19.H), with each load operation targeting four consecutive registers.
Next, the FMOPA instruction takes the first two columns of $Ac$ (Z0.H, Z1.H) as the left-hand operands and performs outer product operations with $Bc$ register pairs (Z4.H, Z5.H) and (Z6.H, Z7.H), accumulating the results into ZA0.S and ZA1.S. The same vector pair is then multiplied with (Z8.H, Z9.H) and (Z10.H, Z11.H) to further update ZA2.S and ZA3.S. 
Similarly, the third and fourth columns of $Ac$ (Z2.H, Z3.H) follow the same pattern with subsequent $Bc$ data, completing a full update of the four ZA tiles.
Similar to the Algorithm~\ref{alg:fp32main_kernel}, we also utilize the Z registers as buffering and employ loop unrolling to improve performance.
After completing the update of $Cr$, the ZA storage results are moved to Z registers in slices and then stored back to $Cr$.
Through this carefully designed register allocation and instruction scheduling, the micro-kernel achieves high instruction-level parallelism and throughput.

\section{Evaluation}
\subsection{Experimental Setup}
\subsubsection{Hardware Platform} 
We evaluate \SystemName  on the Apple M4 Pro chip with SME units. Table~\ref{tab:hardware} details the hardware specifications. M4 Pro employs a shared 16MB L2 cache per cluster of five P-cores and a 4MB L2 cache for all E-cores. Each cluster features a shared SME unit. Experiments are performed on macOS 15.1, with all benchmarks compiled using AppleClang 16.

\begin{table}[ht]
    \centering
    \caption{Main characteristics of Apple M4 Pro.}
    \label{tab:hardware}
    \small
    \begin{tabular}{l l l}
    \toprule
    \textbf{} & \textbf{P-cores} & \textbf{E-cores} \\
    \midrule
    \textbf{Number of Cores}    & 10     & 4 \\
    \textbf{Clock Frequency}    & 4.4 GHz & 2.85 GHz \\
    \textbf{L1 Cache}         & 128 KB & 64 KB \\
    \textbf{L2 Cache}           & 16 MB  & 4 MB \\
    \textbf{Memory Cache}     &      \multicolumn{2}{c}{24 MB} \\
    \textbf{Memory Bandwidth}   & \multicolumn{2}{c}{273 GB/s} \\
    \bottomrule
    \end{tabular}
    \vspace{-2mm}
\end{table}

\subsubsection{Workloads}
To meet practical demands, we benchmark real-world GEMM workloads derived from the DeepSeek~\cite{deepgemm} and LLaMA~\cite{tmac} models. The evaluated workloads span both large and skinny matrices, allowing us to assess performance robustness across diverse aspect ratios. 
Detailed matrix dimensions are summarized in Table~\ref{tab:workloads}.
Following prior work~\cite{libshalom, setting}, all matrices are initialized with random values. We also test irregular matrices with dimensions not divisible by the micro-kernel sizes.

\begin{table}[ht]
\centering
\caption{Configurations of GEMMs in DeepSeek (IDs 1$\sim$18) and LLaMA (IDs 19$\sim$24).}
\label{tab:workloads}
\begin{tabular}{c r r r ||c r r r}
\toprule
\textbf{ID} & \textbf{M} & \textbf{N} & \textbf{K} &
\textbf{ID} & \textbf{M} & \textbf{N} & \textbf{K} \\
\midrule
\textbf{1}  & 64   & 2112  & 7168  & \textbf{2}  & 64   & 24576 & 1536  \\
\textbf{3}  & 64   & 32768 & 512   & \textbf{4}  & 64   & 7168  & 16384 \\
\textbf{5}  & 64   & 4096  & 7168  & \textbf{6}  & 64   & 7168  & 2048  \\
\textbf{7}  & 128  & 2112  & 7168  & \textbf{8}  & 128  & 24576 & 1536  \\
\textbf{9}  & 128  & 32768 & 512   & \textbf{10} & 128  & 7168  & 16384 \\
\textbf{11} & 128  & 4096  & 7168  & \textbf{12} & 128  & 7168  & 2048  \\
\textbf{13} & 4096 & 2112  & 7168  & \textbf{14} & 4096 & 24576 & 1536  \\
\textbf{15} & 4096 & 32768 & 512   & \textbf{16} & 4096 & 7168  & 16384 \\
\textbf{17} & 4096 & 4096  & 7168  & \textbf{18} & 4096 & 7168  & 2048  \\
\textbf{19} & 4096 & 256  & 4096  & \textbf{20} & 11008 & 256  & 4096 \\
\textbf{21} & 4096 & 256  & 11008  & \textbf{22} & 5120 & 256  & 5120 \\
\textbf{23} & 13824 & 256  & 5120  & \textbf{24} & 5120 & 256  & 13824 \\
\bottomrule
\end{tabular}
\vspace{-2mm}
\end{table}

\subsubsection{Baselines}
We compare \SystemName against four libraries that exploit SME to optimize GEMM, including Accelerate, LIBXSMM, KleidiAI and OpenBLAS. Accelerate, LIBXSMM, and KleidiAI support SME2, while OpenBLAS supports SME only. Libraries such as BLIS~\cite{blis} lack SME support and are therefore excluded from comparison.
Accelerate supports transposed inputs $A$ and/or $B$ as well as different storage orders (row-major or column-major) for all three matrices, whereas open-source implementations are incomplete. For example, LIBXSMM only supports the column-major $C$ += $AB$ ($\beta$ = 1) operation, while OpenBLAS and KleidiAI are restricted to the row-major $C$ = $AB$ ($\beta$ = 0) operation. 
All baseline libraries use the officially recommended interfaces and optimal configurations.
\SystemName is configured to align with their computational modes, such as setting $\beta = 1$ for column-major storage.
Each experiment is repeated five times, and the arithmetic mean of the results is reported to obtain stable performance. Warm-up runs are performed for each library to minimize measurement noise.



\subsection{GEMM Performance}
We evaluate FP32 GEMM performance for both row-major and column-major storage on single and multiple SME units.  
This experiment demonstrates the performance gains achieved by the GEMM optimizations in Section~\ref{sec:design}. 

\begin{figure*}[t]
    \centering
    \begin{subfigure}{0.49\textwidth} 
        \centering
    \includegraphics[width=1\linewidth]{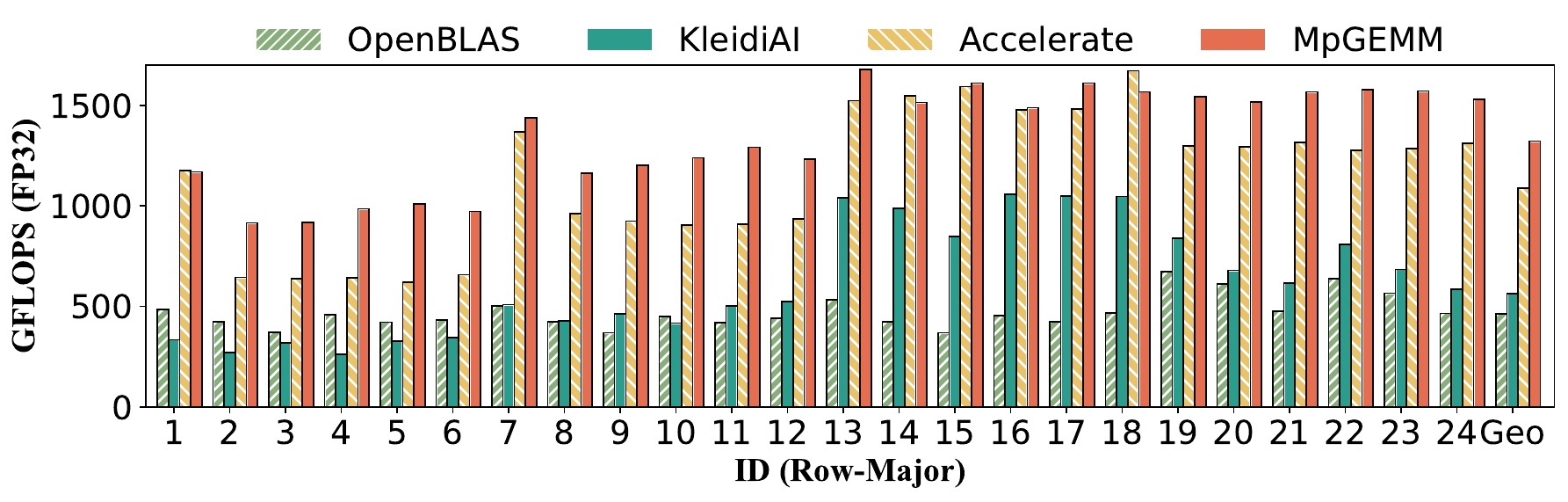}
        \caption{Row-major order storage}
        \label{fig:singlerowmajor} 
    \end{subfigure}
    \hfill
    \begin{subfigure}{0.49\textwidth}
        \centering
    \includegraphics[width=1\linewidth]{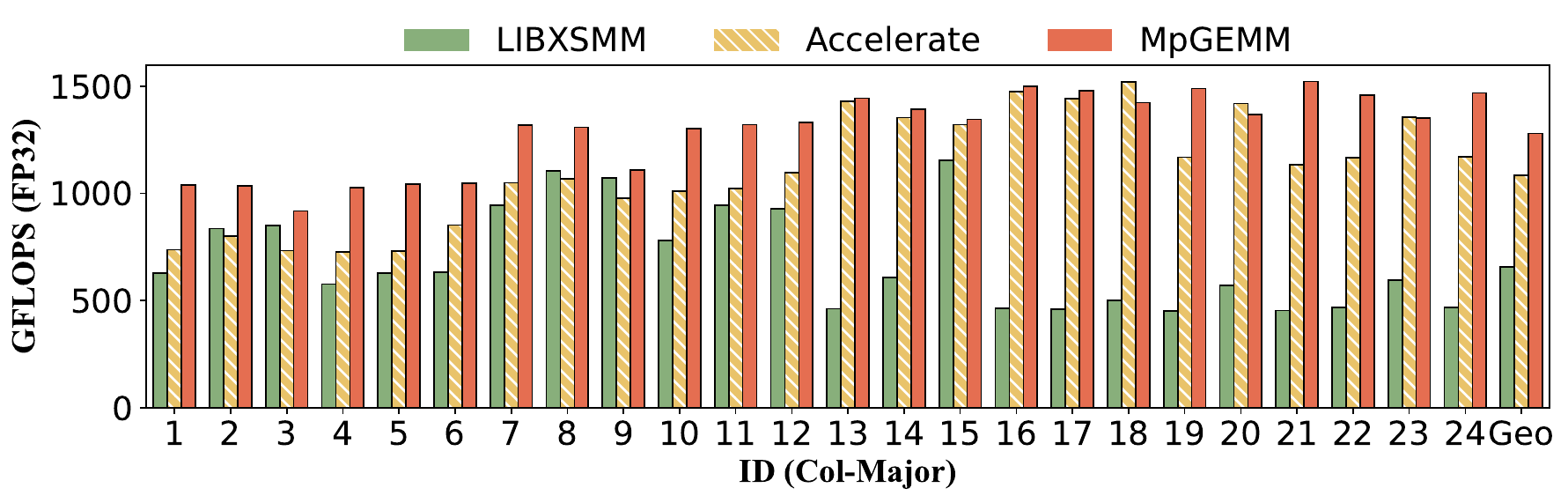}
        \caption{Column-major order storage}
        \label{fig:singlecolmajor} 
    \end{subfigure}
    \caption{The single precision GEMM performance on a single SME unit.}
    \vspace{-3mm}
    \label{fig:singleSME}
\end{figure*}


\begin{figure*}[t]
    \centering
    \begin{subfigure}{0.49\textwidth} 
        \centering
    \includegraphics[width=1\linewidth]{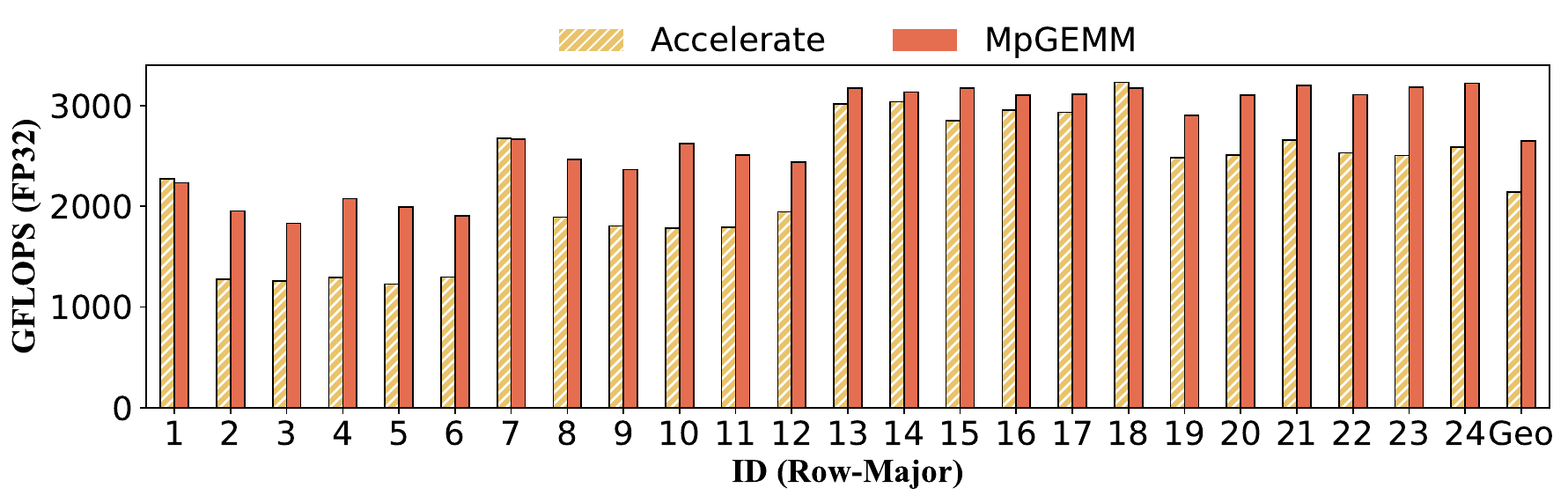}
        \caption{Row-major order storage}
        \label{fig:multirowmajor} 
    \end{subfigure}
    \hfill
    \begin{subfigure}{0.49\textwidth}
        \centering
    \includegraphics[width=1\linewidth]{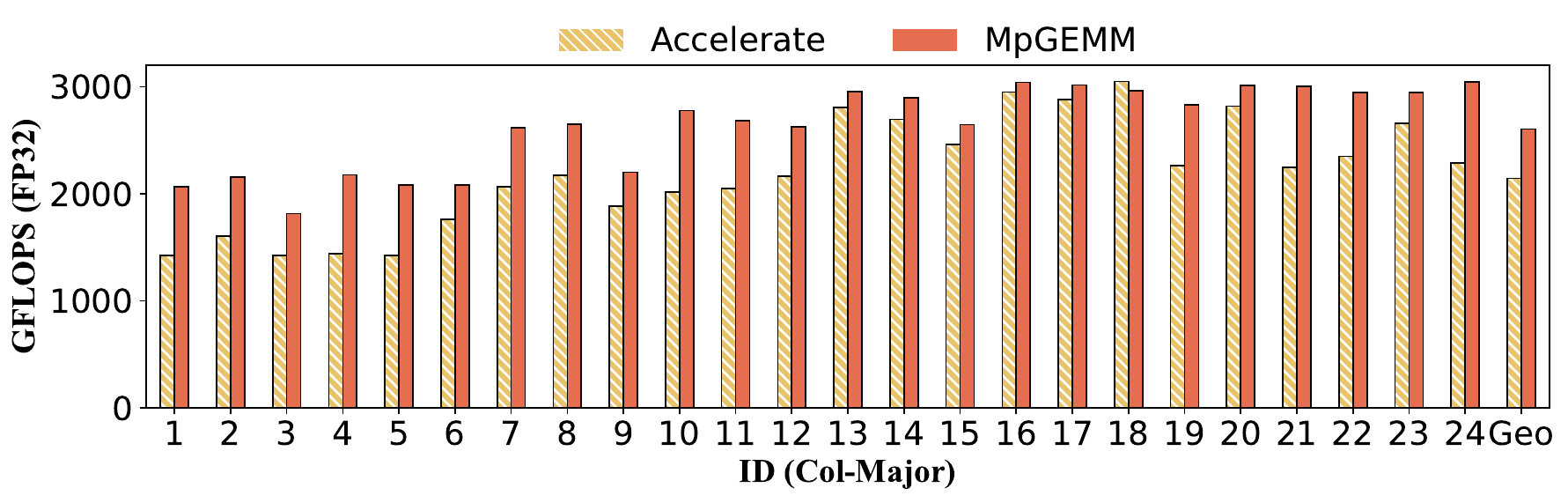}
        \caption{Column-major order storage}
        \label{fig:multicolmajor} 
    \end{subfigure}
    \caption{The single precision GEMM performance on multiple SME units.}
    \vspace{-2mm}
    \label{fig:multipleSME} 
\end{figure*}

\subsubsection{Single SME Unit Performance}
As shown in Fig.~\ref{fig:singleSME},
\SystemName  achieves significant performance advantages over OpenBLAS, KleidiAI, and LIBXSMM, with average speedups of 2.85$\times$, 2.34$\times$, and 1.95$\times$, respectively. The horizontal axis corresponds to the workload IDs listed in Table~\ref{tab:workloads}.

A common limitation of baseline libraries is the inefficient use of the shared L2 cache and L1D TLB. LIBXSMM and OpenBLAS pack only one input matrix to reduce overhead: $B$ is packed in column-major storage, while $A$ is packed in row-major storage, both aligning with the outer product memory layout.
As can be seen in Fig.~\ref{fig:singlecolmajor}, LIBXSMM can surpass Accelerate when the unpacked matrix fits into the L2 cache and the compute-to-memory ratio is high (e.g., IDs 2, 3, 8, and 9). 
Once the working set exceeds the L2 cache, performance degrades due to large-stride and discontiguous accesses that cause frequent TLB and cache misses. In the row-major case (IDs 1-18), matrix $B$ exceeds the L2 cache, while in the column-major case (IDs 13, 14, and 16-24), matrix $A$ does.
A key shortcoming of OpenBLAS is that it does not support SME2 multi-vector data processing instructions, resulting in single-Z-register operations for both computation and memory access.
We should also avoid allocating space for the packed matrix on the stack, as done in LIBXSMM, since this may cause stalls due to memory ordering between the core and SME unit.
KleidiAI pre-packs both input matrices by allocating $M \times K$ and $K \times N$ buffers to store the rearranged $A$ and $B$. This packing overhead results in lower performance than OpenBLAS in small cases (IDs 1-6). However, when $M $ = 4096, the higher compute-to-memory ratio effectively amortizes the packing cost.
To overcome their shortcomings, \SystemName employs an analytical model to guide matrix partitioning, transforming the simple three-nested loop into a six-level blocked structure optimized for the shared L2 cache and TLB. Furthermore, \SystemName improves memory locality by rearranging both input matrices through on-the-fly transposition and first-round online packing.
Another limitation of baseline libraries is the suboptimal multi-vector load method. \SystemName  uses four-way load instructions consistently during both packing and micro-kernel loading.

While Accelerate leverages proprietary kernels, our open-source implementation demonstrates competitive performance under identical workloads, offering a transparent reference for SME-aware optimization. As shown in Fig.~\ref{fig:singleSME}, \SystemName outperforms Accelerate by 1.21$\times$ and 1.18$\times$ on average for row-major and column-major layouts, respectively.


\begin{figure}[t]
\centerline{\includegraphics[width=1\linewidth]{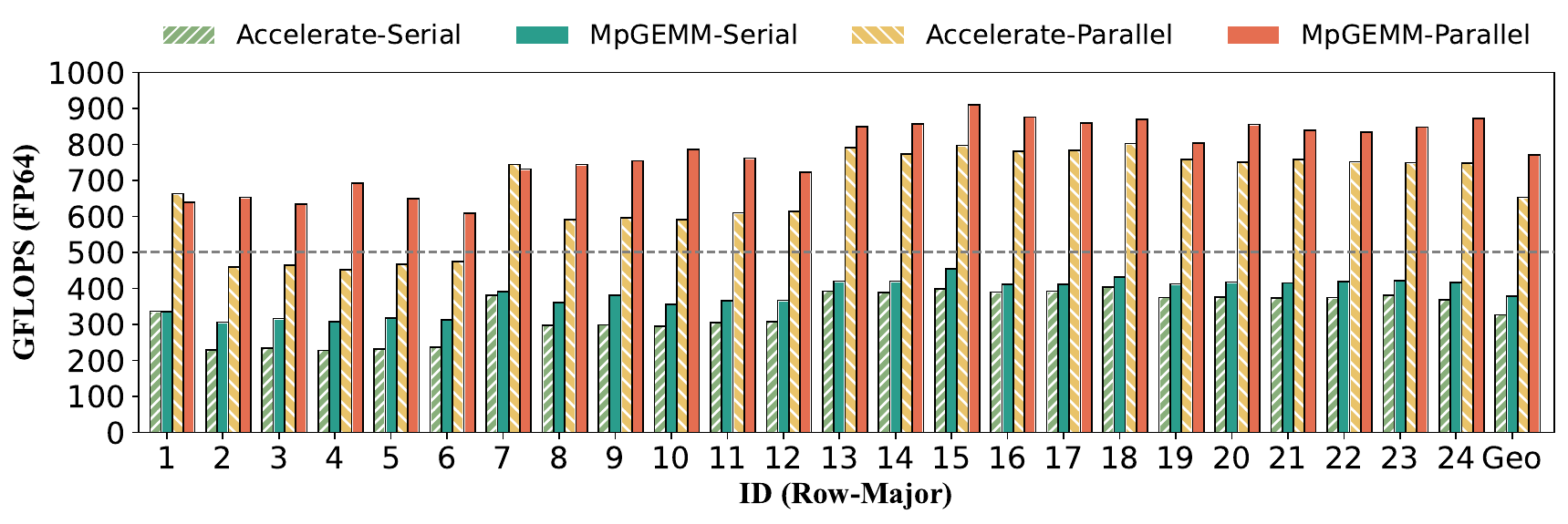}
}
\caption{The double precision GEMM performance on single and multiple SME units.}
\label{fig:row-fp64} 
\vspace{-2mm}
\end{figure}

\begin{figure}[t]
\centerline{\includegraphics[width=1\linewidth]{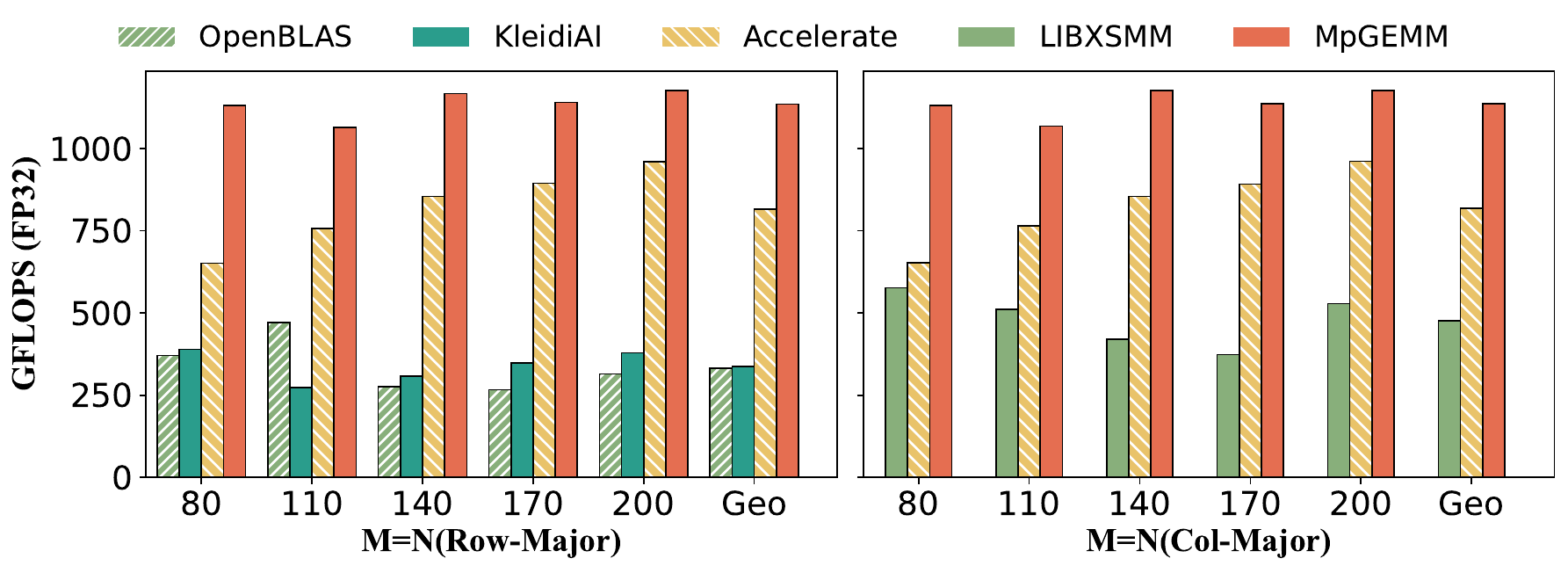}
}
\caption{The performance of irregular-shaped GEMM on a single SME unit.}
\label{fig:irregular} 
\vspace{-4mm}
\end{figure}

\begin{figure*}[t] 
  \centering
  \begin{subfigure}[b]{0.49\textwidth}
    \centering
    \includegraphics[width=1\linewidth]{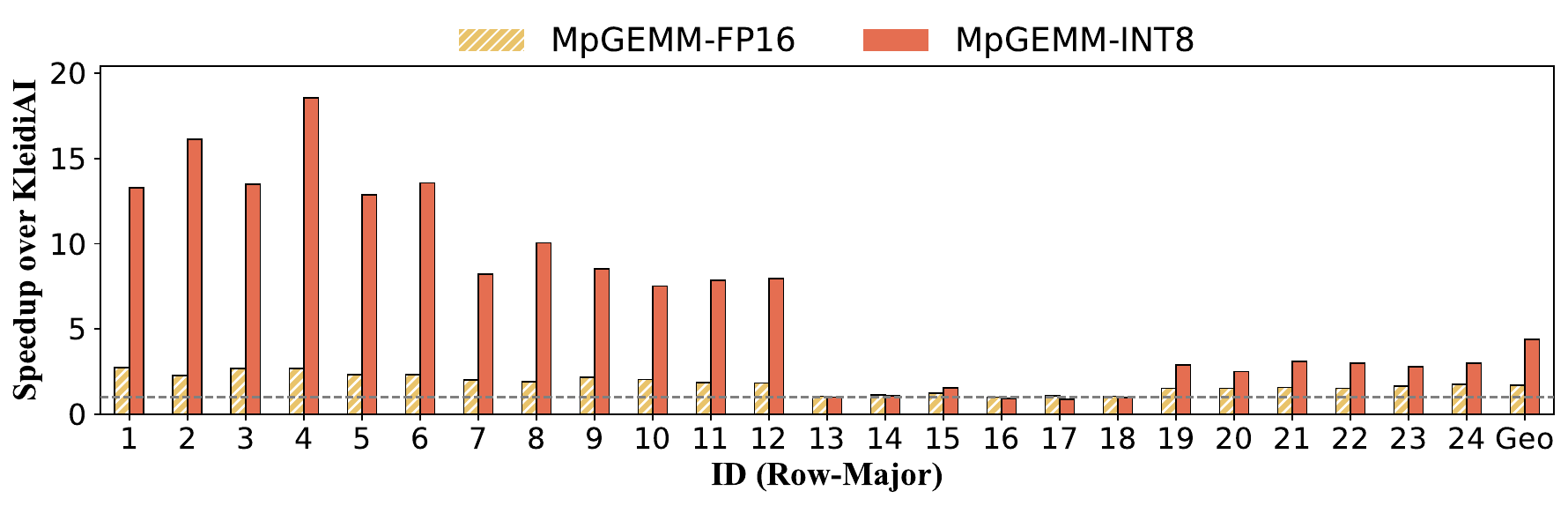}
    \caption{The mixed-precision GEMM performance on a single SME unit.}
    \label{fig:fp16gemm}
  \end{subfigure}
  \hfill
  \begin{subfigure}[b]{0.49\textwidth}
    \centering
    \includegraphics[width=1\linewidth]{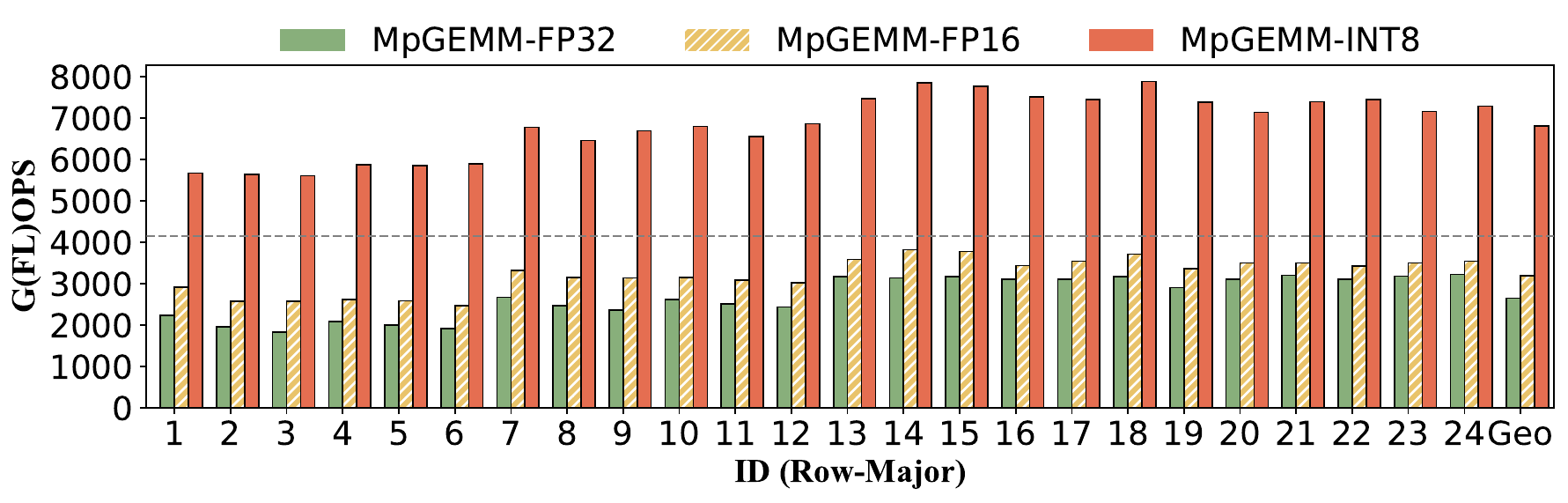}
    \caption{Performance comparison of \SystemName  on multiple SME units.}
    \label{fig:fp32-fp16}
  \end{subfigure}
  \caption{Performance of GEMM under different precisions (Row-major).}
  \label{fig:gemm-comparison}
  \vspace{-4mm}
\end{figure*}

\subsubsection{Multiple SME Unit Performance}
Due to the limited number of SME units on the Apple M4 Pro platform, we focus on evaluating parallel efficiency instead of scalability. 
To avoid performance degradation caused by multiple threads in a cluster accessing the shared SME unit simultaneously, we limit the number of threads per cluster to two. 
\SystemName achieves a twofold performance improvement by dynamically distributing computational tasks among all available SME units.
Among existing libraries, only Accelerate supports parallel matrix multiplication across multiple SME units. Fig.~\ref{fig:multipleSME} compares the multi-core performance of \SystemName and Accelerate, with experimental settings identical to the single-core evaluation. 
The results show that \SystemName achieves average single precision GEMM speedups of 1.24$\times$ and 1.22$\times$ over Accelerate for row-major and column-major layouts, respectively.
Compared to LIBXSMM, KleidiAI, and OpenBLAS, which fail to implement SME-based parallelism, \SystemName achieves average speedups of 3.96$\times$, 4.69$\times$, and 5.7$\times$, respectively. 

In addition, we evaluate double precision GEMM performance for the row-major layout against Accelerate, the only library supporting FP64 GEMM on SME (Fig.~\ref{fig:row-fp64}).
Despite the architectural limitation that FP64 throughput is only one-quarter of FP32, \SystemName achieves a speedup of 1.18$\times$ under multi-core execution.

\subsubsection{Irregular Matrices}
When the matrix size is not a multiple of the micro-kernel size, we need to use predication registers to mask trailing elements. This limits the utilization of ZA tiles and impedes overall performance.
To evaluate the boundary handling capabilities of different GEMM implementations, we vary $M$ and $N$ from 80 to 200 in steps of 30 to avoid matrix sizes being integer multiples of a single ZA tile. Like prior works~\cite{libshalom,dsp}, $K$ is set to a large value (25600) to prevent performance bias caused by hot data in caches. 
In Fig.~\ref{fig:irregular}, LIBXSMM outperforms OpenBLAS and KleidiAI because it provides edge $16\times64$ and $64\times16$ micro-kernels for boundary handling, enabling full utilization of all ZA tiles.
With efficient packing and optimized main and edge micro-kernels, \SystemName  consistently outperforms alternatives, even on irregular dimensions.

\subsection{Mixed-Precision GEMM Performance}
In this experiment, we evaluate the throughput of mixed-precision on both single and multiple SME units, considering FP16 and INT8 inputs accumulated to FP32 and INT32, respectively. We compare \SystemName  with KleidiAI to verify the effectiveness of the optimization strategies proposed in Section~\ref{sec:mpdesign}. KleidiAI supports BF16 instead of FP16, with its BFMOPA achieving the same single-core peak performance (2006 GFLOPS) as FP16 FMOPA. 
In all tests, our GEMM with FP16 inputs implementation consistently outperforms KleidiAI, achieving an average speedup of 1.7$\times$.
This improvement stems from efficient cache utilization through refined partitioning and  high-bandwidth four-way load.
The SMOPA instruction performs the outer product of two INT8 vectors and accumulates the results into INT32, providing up to 4010 GOPS of computational throughput on a performance SME unit. Therefore, our INT8 GEMM achieves an average 2$\times$ speedup over the FP16 GEMM baseline. In contrast, KleidiAI implements INT8 GEMM using ARM NEON for data packing. For flat GEMMs, however, the limited reuse of packed data prevents the packing overhead from being amortized, thereby constraining overall performance.
SME natively supports FP16/INT8 types, allowing FP16/INT8 inputs to be multiplied and accumulated into FP32/INT32 outputs. 
Higher-precision accumulation reduces precision loss during matrix multiplication,
making it well suited for deep learning workloads~\cite{fathom}.



Fig.~\ref{fig:fp32-fp16} shows the multi-core performance of \SystemName across three input precisions: FP32, FP16, and INT8. Microbenchmarks indicate that the throughput of FP16 FMOPA is equivalent to that of FP32 FMOPA. However, the mixed-precision GEMM with FP16 inputs achieves better practical performance because of its higher compute-to-memory ratio. The INT8 GEMM achieves 7856 and 7882 GOPS in the IDs (14, 18) tests, respectively, which corresponds to about 94\% of the SMOPA peak performance measured on SME.



\begin{figure}[t]
\centerline{\includegraphics[width=1\linewidth]{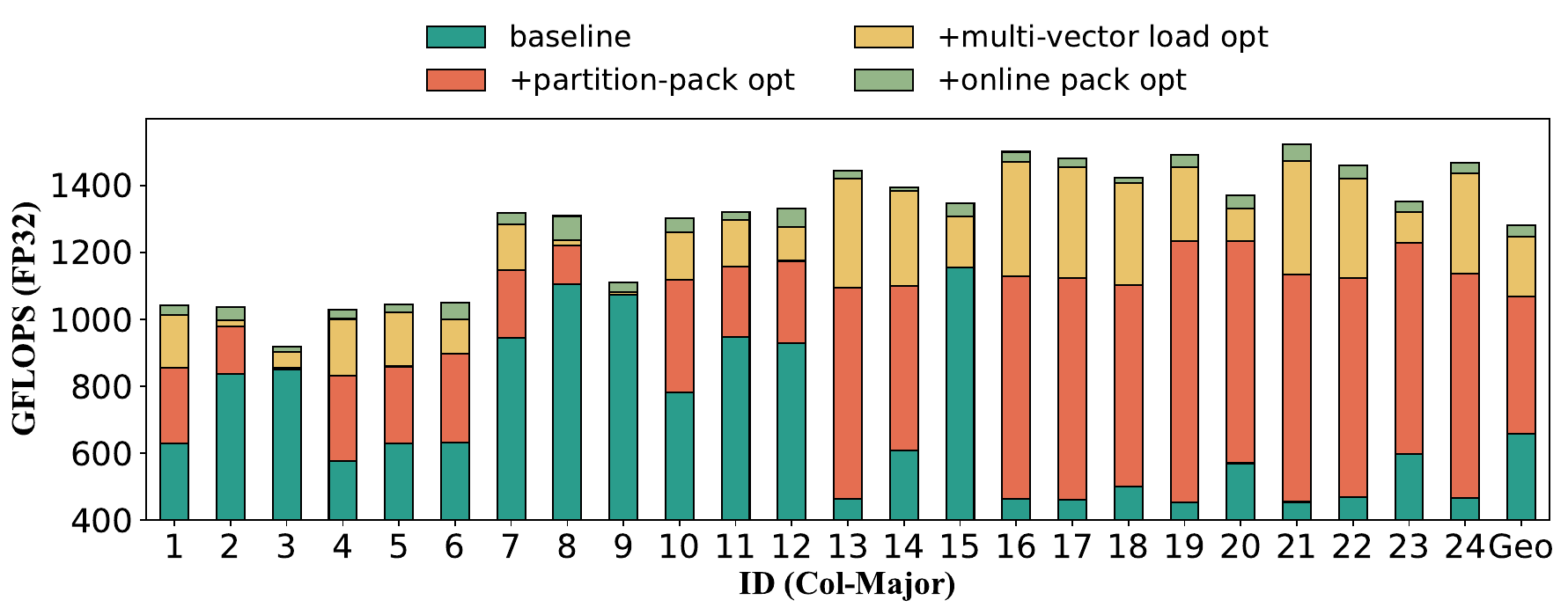}
}
\caption{Breakdown of Optimizations on a single SME unit.}
\label{fig:optbreakdow} 
\vspace{-4mm}
\end{figure}
\subsection{Optimization Breakdown}
To assess the effectiveness of the optimization methods presented in Section~\ref{sec:design}, we evaluate how three strategies, cache-aware partitioning and dual-matrix packing, four-way loading, and first-round online packing, contribute to performance improvement. In this experiment, LIBXSMM is used as a baseline to quantify the performance gains achieved by these optimizations.  
As shown in Fig.~\ref{fig:optbreakdow}, our partitioning and packing schemes and multi-vector load method deliver significant performance improvements, achieving average speedups of 1.62$\times$ and 1.17$\times$, respectively. 
The former transforms the naive three-loop structure into a six-level blocked algorithm and rearranges blocks to efficiently leverage vector load and outer product instructions. This process involves fine-tuning the tiling parameters $mc$, $nc$, and $kc$ to ensure that sub-blocks remain resident in the shared L2 cache. The latter reconstructs the micro-kernel to maintain four-way loading while covering both packing and micro-kernel stages. In contrast, the benefit of first-round online packing is limited, as FMOPA cycles are insufficient to fully hide memory write overhead.
In the IDs (3, 9, 15) tests ($K$ = 512), we observe that the performance obtained from partitioning and packing operations on both matrices is lower than that of LIBXSMM.
\SystemName  still achieves better performance by exploiting four-way load and online pack optimizations.

\section{Related Works}
Optimizing GEMM has long been a central topic in HPC. Goto et al.~\cite{goto} introduced multi-level blocking and packing strategies, laying the foundation for high-performance BLAS libraries such as OpenBLAS~\cite{openblas}. LIBXSMM~\cite{libxsmm} further extends this approach by generating JIT-compiled micro-kernels for small matrices. However, mainstream linear algebra libraries still deliver suboptimal performance on the ARM SME architecture, leaving much room for improvement.

Hardware vendors have introduced dedicated matrix computation units, such as NVIDIA Tensor Cores~\cite{tensorcore}, Google TPUs~\cite{tpu}, and Intel AMXs~\cite{amx}, accompanied by highly optimized software libraries. These accelerators operate on fixed-size matrix tiles and rely on compiler or runtime support to achieve peak performance. In contrast, SME supports flexible vector outer product instructions with scalable vector lengths, requiring tailored optimizations.

Several efforts have focused on optimizing GEMM on ARM SME CPUs. Accelerate framework has integrated SME units to accelerate FP32 and FP64 GEMMs. However, it is proprietary and confined to Apple platforms, thereby limiting both portability and transparency.
Remke et al.~\cite{hellosme} conducted the first evaluation of memory bandwidth and computational instructions on the M4 processor with SME units. They developed fast kernels tailored for small-scale GEMMs and integrated them into LIBXSMM. OpenBLAS adopts a method similar to LIBXSMM, without support for vector-group data processing instructions on SME2.
KleidiAI~\cite{kleidiai} leverages NEON MLA and SME2 outer product instructions to accelerate basic DL kernels, but does not provide any memory management functionality.
These solutions fail to exploit the SME memory hierarchy and the high-bandwidth capabilities of multi-vector load operations, resulting in poor performance on large-scale GEMMs.
Different from prior works, \SystemName is the first library to systematically exploit SME architectural features for GEMM optimization across multiple precisions. 
It introduces a cache-aware algorithm for GEMM to maximize spatial locality after packing. \SystemName also employs efficient dual packing along with dedicated micro-kernels, each using four-Z-register grouped loads. Experimental results show that our approach significantly improves performance.

\section{Conclusion}
This paper introduces \SystemName, an open-source library optimizing multiple precision GEMM performance on ARM SME CPUs. 
We characterize SME's architectural features, including tile reuse, load granularity, and its compute and memory hierarchies, which guide GEMM optimization.
\SystemName employs a cache-aware partitioning strategy guided by an analytical model, along with dual-matrix packing schemes to enhance memory access efficiency and cache utilization.
To mitigate the overhead of packing, we introduce on-the-fly transposition and first-round online packing techniques.
We also design dedicated main and edge micro-kernels to exploit multi-vector loads and maximize tile register reuse.
These optimizations are further generalized to mixed-precision GEMMs.
Experimental results demonstrate that \SystemName achieves an average speedup of 1.23$\times$ over the vendor-optimized BLAS library in Accelerate and significantly
outperforms other open-source solutions.



\bibliographystyle{IEEEtran}
\balance
\bibliography{refs}

\end{document}